\documentclass[sigconf,balance=false]{acmart}
\usepackage{popets}
\setcopyright{popets}
\copyrightyear{YYYY}
\acmYear{YYYY}
\acmVolume{YYYY}
\acmNumber{X}
\acmDOI{XXXXXXX.XXXXXXX}
\acmISBN{}
\acmConference{Proceedings on Privacy Enhancing Technologies}
\usepackage{algorithm}
\usepackage{algorithmic}
\usepackage{enumitem}
\usepackage{graphicx}
\usepackage{multirow}
\usepackage{booktabs}
\usepackage{subcaption}
\usepackage{caption}
\usepackage{titlesec}
\usepackage{xspace}
\usepackage{amsmath}
\usepackage{threeparttable}

\expandafter\def\expandafter\normalsize\expandafter{%
    \normalsize%
    \setlength\abovedisplayskip{3pt}%
    \setlength\belowdisplayskip{3pt}%
    \setlength\abovedisplayshortskip{2pt}%
    \setlength\belowdisplayshortskip{2pt}%
}

\titlespacing*{\subsection}{0pt}{*0.2}{*0.2}
\titlespacing*{\section}{0pt}{*0.4}{*0.4}

\newcommand{\etal}{{\em et al.}\xspace}
\newcommand{\BfPara}[1]{{\noindent {\bf #1.}}}

\begin{document}

\title{Unveiling Client Privacy Leakage from Public Dataset Usage in Federated Distillation}

\author{Haonan Shi}
\email{haonan.shi3@case.edu}
\affiliation{%
  \institution{Case Western Reserve University}
  \city{Cleveland}
  \state{Ohio}
  \country{USA}
}

\author{Tu Ouyang}
\email{tu.ouyang@case.edu}
\affiliation{%
  \institution{Case Western Reserve University}
  \city{Cleveland}
  \state{Ohio}
  \country{USA}
}

\author{An Wang}
\email{an.wang@case.edu}
\affiliation{%
  \institution{Case Western Reserve University}
  \city{Cleveland}
  \state{Ohio}
  \country{USA}
}


\begin{abstract}
Federated Distillation (FD) has emerged as a popular federated training framework, enabling clients to collaboratively train models without sharing private data.
Public Dataset-Assisted Federated Distillation (PDA-FD), which leverages public datasets for knowledge sharing, has become widely adopted. 
Although PDA-FD enhances privacy compared to traditional Federated Learning, we demonstrate that the use of public datasets still poses significant privacy risks to clients' private training data.
This paper presents the first comprehensive privacy analysis of PDA-FD in the presence of an honest-but-curious server. 
We show that the server can exploit clients' inference results on public datasets to extract two critical types of private information: label distributions and membership information of the private training dataset.
To quantify these vulnerabilities, we introduce two novel attacks specifically designed for the PDA-FD setting: a label distribution inference attack and innovative membership inference methods based on Likelihood Ratio Attack (LiRA).
Through extensive evaluation of three representative PDA-FD frameworks (FedMD, DS-FL, and Cronus), our attacks achieve state-of-the-art performance, with label distribution attacks reaching minimal KL-divergence and membership inference attacks maintaining high True Positive Rates under low False Positive Rate constraints. 
Our findings reveal significant privacy risks in current PDA-FD frameworks and emphasize the need for more robust privacy protection mechanisms in collaborative learning systems.
\end{abstract}


\maketitle

\section{Introduction}
In recent years, federated learning (FL) has emerged as a promising paradigm for collaborative machine learning while preserving data privacy~\cite{mcmahan2017communication}.
Traditional FL frameworks, such as FedAvg~\cite{mcmahan2017communication} and FedSGD, require clients to upload model parameters or gradients to a central server for aggregation, which can introduce limitations in both privacy and utility.
To address these issues, federated distillation (FD)~\cite{li2019fedmd, sui2020feded, itahara2021distillation, chang2019cronus, jeong2018communication} has gained attention as an alternative approach that offers enhanced privacy protection and reduced communication overhead~\cite{wu2022communication,huang2023decentralized}.
In FD, model-inference outputs or distilled knowledge are exchanged between the server and clients instead of model parameters.
This learning scheme only requires black-box access to client models, supporting diverse model architectures across clients.
Existing approaches, such as FedMD~\cite{li2019fedmd}, DS-FL~\cite{itahara2021distillation} and Cronus~\cite{chang2019cronus}, have been proposed to further enhance the privacy protection and efficiency of collaborative learning.

\begin{figure}
    \centering
    \includegraphics[width=1\linewidth]{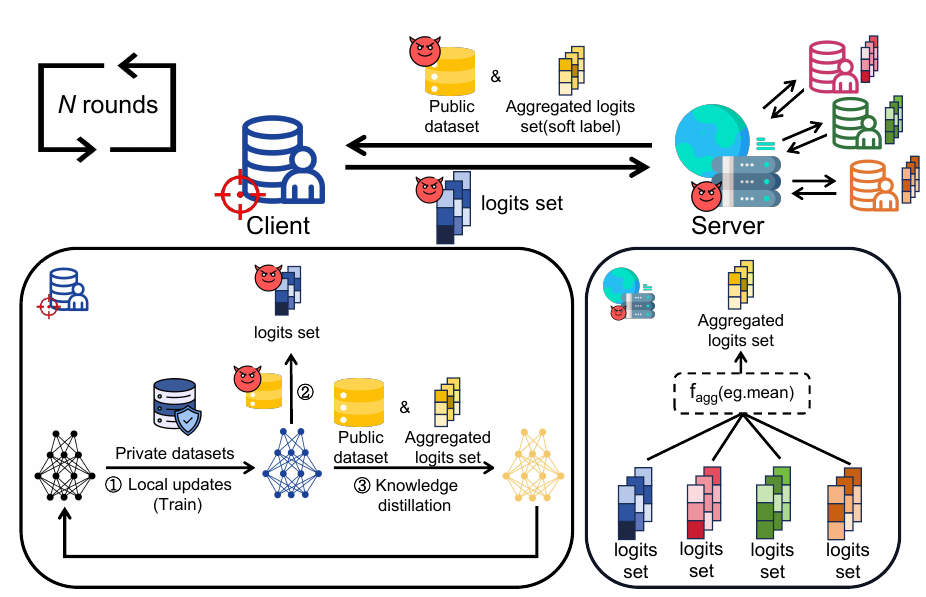}
    \caption{Workflow of Public Dataset-Assisted Federated Distillation (PDA-FD). During each collaborative training round, clients first train their private models on their respective private datasets, then perform inference on the public dataset and transmit the resulting logits back to the server as transferred knowledge. As an \textit{honest-but-curious} server, the server can extract private information from target clients' private datasets by manipulating and leveraging public datasets.}
    \label{fig:FD_workflow}
\end{figure}

In these solutions, public datasets are often used to facilitate knowledge distillation among clients.
Knowledge sharing can be achieved across clients with diverse data distributions by having all clients perform inference on the same public data and sharing these inference results as knowledge.
We call such a learning scheme public dataset-assisted federated distillation, or PDA-FD.
Despite the benefits of FD, the privacy implications in such frameworks have not been thoroughly explored in the existing literature.
While FD generally provides stronger privacy guarantees than traditional FL, using a public dataset for information exchange can result in potential privacy leakage. 
Figure \ref{fig:FD_workflow} illustrates the workflow of PDA-FD. In each collaborative training round of PDA-FD, clients need to train their private models on their respective private datasets before performing inference on the public dataset and transmitting the inference results as knowledge to the server. This process enables better transfer of knowledge learned from private data to other clients' private models. However, this process also enhances the memorization of private datasets by private models, thereby causing the private models' inference results on the public dataset to leak private information from the private dataset. 
We discover that an \textit{honest-but-curious} server, via manipulating the public dataset and exploiting the client' inference results on the public dataset, can obtain private information from a particular client's private training dataset without compromising the training process of PDA-FD. 
In particular, the label distribution of the training dataset, and if specific data belongs to a client's training dataset (membership information). These two information leaks are representative and frequently studied in the ML privacy literature~\cite{gu2023ldia, yang2022fd, nasr2019comprehensive}.

In this work, we devise two types of privacy attacks against clients by the server: \textit{Label Distribution Inference Attacks (LDIA)} and \textit{Membership Inference Attacks (MIA)}.
LDIA reveals privacy information about the overall data distribution of all client private datasets combined. In contrast, membership inference attacks enable more granular privacy leakage by identifying the presence of specific data samples in a client's private datasets.
Both LDIA and MIA can be performed in a black-box manner, requiring only clients' inference results on the public dataset, which renders these attacks practically useful since FD frameworks are generally designed to not transfer a client model to the server.
Additionally, Federated Distillation assumes non-IID data distributions across clients in the label space.
LDIA becomes particularly important in this context, as inferring the label distribution can reveal significant information about a client's unique data characteristics. 
For example, if hospitals were conducting FD to protect patient privacy, attackers could expose which facilities specialize in rare conditions, potentially de-anonymizing patients with unique medical profiles while also compromising institutional privacy.
Furthermore, LDIA and MIA can serve as stepping stones for more sophisticated attacks.
For instance, the obtained knowledge can be leveraged to generate synthetic data that mimics private datasets.

In the recent literature, LDIA and MIA have been extensively studied in the traditional FL and centralized machine learning settings~\cite{nasr2019comprehensive,shokri2017membership,yeom2018privacy,melis2019exploiting, ramakrishna2022inferring, jiang2024protecting}.
They have largely focused on white-box or gradient-based attacks.
For example, Gu \etal demonstrated that a malicious server in FL can infer label distributions by exploiting the gradients or parameters uploaded by clients~\cite{gu2023ldia}.
Similarly, Wainakh \etal showed that user-level label leakage is possible through gradient analysis~\cite{wainakh2021user}.
For MIAs, Nasr \etal developed sophisticated MIA techniques that exploit the white-box access to model parameters in FL~\cite{nasr2019comprehensive}.
The rich information leveraged in these attacks are not directly accessible in FD settings.

In the FD setting, a few studies have primarily focused on MIAs, such as \textit{FD-Leaks}~\cite{yang2022fd}, \textit{MIA-FedDL}~\cite{liu2023mia} and \textit{GradDiff}~\cite{wang2024graddiff}.
However, due to reasons such as the use of low-quality shadow models or overly strong threat model assumptions, these methods tend to experience decreased effectiveness in real-world FD scenarios.
In \textit{FD-leaks}, a malicious client leverages its local model as a shadow model to train an MIA classifier, then applies the classifier on target sample inference score to determine target sample membership in other clients' private datasets.
Effective MIA requires attackers to possess high-quality shadow models, which are shadow models that can mimic target model performance. Such shadow models need to be trained on shadow datasets sharing similar data distributions with the target model's training dataset~\cite{shokri2017membership, carlini2022membership}.
However, in \textit{FD-leaks}, the distribution of the training dataset of the malicious client's local model may not be consistent with the distribution of the target model's training dataset, resulting in the shadow model used by the malicious client potentially being a low-quality shadow model, which leads to limited MIA performance.
In \textit{MIA-FedDL}~\cite{liu2023mia}, the attacker acts as clients, while in \textit{GradDiff}~\cite{wang2024graddiff}, the attacker can be either clients or the server. 
However, for \textit{MIA-FedDL} and \textit{GradDiff}, they assume that the attacker can additionally obtain a carefully designed shadow dataset with a distribution consistent with the target client model's training dataset distribution to train the shadow models used in MIA. 
We consider this to be an overly strong threat model assumption in the FD scenario.
In the FD setting, it is challenging for both clients and the server to obtain the data distribution of other clients' private data. 
Therefore, in a more realistic threat model where clients and servers cannot directly obtain the private data distribution of the target client, 
these MIA methods~\cite{yang2022fd, liu2023mia, wang2024graddiff} in heterogeneous non-IID environments will experience decreased effectiveness due to the inability to use high-quality shadow models in MIA.
Simultaneously, we note that these MIAs in FD scenarios still combine the use of low-precision classifier-based MIA~\cite{shokri2017membership}, resulting in their relatively limited performance on the key evaluation metric of True Positive Rate at low False Positive Rate Region~\cite{carlini2022membership}.

To address the limitations of the existing works,
we aim to comprehensively and effectively examine privacy leakage by public datasets in FD across multiple frameworks (FedMD, DS-FL, and Cronus) and various data distribution scenarios.
We introduce new attack methods that are specifically tailored to the PDA-FD setting.
Specifically, we propose a novel LDIA method based on public datasets and extend the state-of-the-art MIA, Likelihood Ratio Attack (LiRA)~\cite{carlini2022membership}, to overcome the challenges posed by the limited information available in PDA-FD.
To that end, we design and implement \textit{Co-op LiRA} and \textit{Distillation-based LiRA} for MIAs.
These two MIA approaches relax the constraint present in most current MIAs~\cite{carlini2022membership, watson2021importance, shokri2017membership} that require the attacker to possess a shadow dataset with the same data distribution as the target model's training dataset.
In order to obtain shadow models that can effectively mimic the target model's performance, in \textit{co-op LiRA}, the server executes LDIA as a preliminary attack to obtain the label distribution of all clients' private data, then selects client models with label distributions similar to the target client model's as shadow models in MIA. In \textit{distillation-based LiRA}, the server utilizes the public dataset used in the FD training process, using the target client model as a teacher model to distill several student models as shadow models in MIA.

Compared to previous MIA methods in FD, for \textit{FD-leaks}~\cite{yang2022fd}, through our approach of selecting or training shadow models, we are able to obtain higher-quality shadow models that can better mimic the performance of target client models. 
Unlike \textit{MIA-FedDL}~\cite{liu2023mia} and \textit{GradDiff}~\cite{wang2024graddiff}, our attack operates under more realistic threat model assumptions in FD scenarios, requiring only the existing public dataset from the FD training process rather than an additional carefully crafted shadow dataset that matches the target model's training distribution.
Furthermore, our proposed MIAs adapt and integrate the advanced MIA method LiRA into realistic FD scenarios with constrained threat model, offering significantly better performance than classifier-based MIAs~\cite{shokri2017membership} used in prior works~\cite{yang2022fd, liu2023mia, wang2024graddiff}, thereby bringing privacy research in FD closer in line with the state-of-the-art developments in the MIA field~\cite{carlini2022membership, watson2021importance}.
Additionally, while some previous works~\cite{yang2022fd,wang2024graddiff} primarily focused on client-side attacks, our work provides a more holistic view by examining both LDIA and MIA from the server's perspective.
A high-level comparison with the existing work is shown in Table~\ref{tab:mia_discussion}.

\begin{table}[h]
    \caption{Summary of Privacy Attacks in FL \& FD}
    \centering
    \begin{threeparttable}
    \scriptsize
    \resizebox{1\linewidth}{!}{%
    \begin{tabular}{c|c|c|c|c}
        \toprule
        Method & Attacker & Framework & Ideal $D_{shadow}$\tnote{*} &Attack Goal\\
        \midrule
        \cite{gu2023ldia}     & Server & FL & Required & LDIA  \\
        \cite{nasr2019comprehensive} & Server\&Client & FL & Required & MIA \\
        \cite{liu2023mia}    & Client & FD & Required & MIA \\
        \cite{yang2022fd}& Client & FD & Not Required & MIA\\
        \cite{wang2024graddiff}& Server\&Client & FD & Required & MIA\\
         \midrule
         Ours & Server & FD & Not Required & LDIA\&MIA \\
        \bottomrule
    \end{tabular}%
    }
    \begin{tablenotes}
    \scriptsize
    \item[*] $D_{shadow}$ that has the same data distribution as $D_{train}$.
    \end{tablenotes}
    \end{threeparttable}
    \label{tab:mia_discussion}
\end{table}


In our study, our key findings include:
(1) LDIA can be successfully achieved across multiple PDA-FD frameworks that significantly outperform random guessing baselines.
(2) The proposed \textit{co-op LiRA} and \textit{distillation-based LiRA} shows high effectiveness in terms of the True Positive Rate (TPR) in a low False Positive Rate (FPR) region.
(3) We also show that the effectiveness of LDIA and MIA varies with data distributions. Adversaries can generally achieve higher attack success rates when data follows more uniform distributions compared to non-IID settings.
Our findings collectively reveal significant privacy risks in current PDA-FD frameworks and highlight the need for more advanced privacy-preserving mechanisms.

\section{Understanding the Privacy Risks in PDA-FD}
\label{sec:background}
\subsection{PDA-FD}
\label{sec:background_fd}

Federated Distillation (FD)~\cite{li2019fedmd, itahara2021distillation, chang2019cronus, jeong2018communication, sui2020feded} is a specialized FL framework distinct from traditional FL. 
FD exchanges model outputs or distilled knowledge between the server and clients instead of model parameters, which significantly reduces communication overhead.
Additionally, FD only requires black-box access to client models and supports diverse model architectures across clients. 
As a result, FD not only better preserves privacy but also offers greater utility compared to traditional FL frameworks.

In our study, we focus on one category of FD frameworks, \textbf{Public Dataset-Assisted Federated Distillation (PDA-FD)}~\cite{li2019fedmd, itahara2021distillation, chang2019cronus}.
The workflow of PDA-FD is shown in Figure \ref{fig:FD_workflow}.
In PDA-FD, the server leverages a public dataset to facilitate knowledge transfer among clients. 
The public dataset is prepared by the server and shared with all the clients during the collaborative training rounds~\cite{li2019fedmd, chang2019cronus, itahara2021distillation}.
The PDA-FD framework typically involves three phases in each collaborative training round: local updates phase, communication phase, and knowledge distillation phase.

During the local updates phase, client $n$ trains its local model $\theta_{n}$ on its private dataset $D_{n}$ using stochastic gradient descent~\cite{lecun1998gradient}. 
The loss function $\mathcal{L}(x, y, \theta_{n})$ is defined to calculate the error between the prediction posterior $f_{\theta_{n}}(x)_y$ of the training data and its ground truth label $y$. 
Cross-entropy is often used as the loss function:
$\mathcal{L}(x, y, \theta_{n}) = -log(f_{\theta_{n}}(x)_y)$.

During the communication phase, a set of data samples $S_t$ are selected by the server from the public dataset. 
For each selected sample $x_k$, the client’s local model $\theta_{n}$ performs inference to obtain the corresponding logits $z_{\theta_{n}}(x_k)$ and sends the logits to the server. 
After collecting the logits from all clients, the server aggregates them using an aggregation algorithm $f_{agg}$(eg. average aggregation~\cite{li2019fedmd}) to get an aggregated logits $Z_k$. 
The server then distributes $Z_k$ back to each client.
At the end of the communication phase, each client will receive the logits set $\{ Z_k \mid k \in S_t \}$ from server.

During the knowledge distillation phase, client $n$ performs knowledge distillation~\cite{hinton2015distilling} using the aggregated logits $Z_k$ returned by the server as the soft labels. 
These aggregated logits $Z_k$ represent knowledge learned by other clients' private models on their own private datasets, allowing clients to learn from other clients' knowledge and further improve their own model's performance.
In this case, the loss function will be the mean absolute errors (MAE):
\begin{equation}
\mathcal{L}(x, y, \theta_{n}) = \frac{1}{N} \sum_{n=1}^{N} \left| Z_k - z_{\theta_n}(x_k) \right|
\label{eq:loss_function_mae}
\end{equation}
or Kullback-Leibler (KL) divergence values~\cite{kullback1951information}:
\begin{equation}
\mathcal{L}(x, y, \theta_{n}) = \sum_{n=1}^{N} Z_k \cdot \log \left( \frac{Z_k}{z_{\theta_n}(x_k)} \right)
\label{eq:loss_function_kl}
\end{equation}
The overall procedure of PDA-FD learning is summarized in Algorithm \ref{alg:FD_algorithm}. Different PDA-FD frameworks~\cite{chang2019cronus, li2019fedmd, itahara2021distillation} require clients to upload either logit vectors or prediction probability vectors during communication.
\begin{algorithm}
\caption{Public Dataset-Assisted Federated Distillation}
\label{alg:FD_algorithm}
\begin{algorithmic}[1]
    \REQUIRE Private datasets $\{D_n\}_{n=1}^N$, public dataset $D_{pub}$, local models $\{\theta_n\}_{n=1}^N$, number of collaborative training round $T$, public data index set $\{S_t\}_{t=1}^T$.
    \FOR{collaborative training round $t = 0$ to $T$} 
       \STATE {\color{blue}\textit{$\triangleright$ Local Updates Phase}}
       \STATE Each client trains local model $\theta_{n}$ on private dataset $D_{n}$
       \STATE {\color{blue}\textit{$\triangleright$ Communication Phase}}
       \STATE Each client computes logits $\{z_{\theta_n}(x_k) \mid k \in S_t\}$ for public data $\{x_k|k \in S_t\}$
       \STATE Each client sends logits set $\{z_{\theta_n}(x_k) \mid k \in S_t\}$ to server
       \FOR{$k \in S_t$}
           \STATE $Z_k \leftarrow f_{agg}(\{z_{\theta_n}(x_k)\}_{n=1}^N)$
       \ENDFOR
       \STATE Server sends aggregated logits $\{Z_k \mid k \in S_t\}$ to clients
       \STATE {\color{blue}\textit{$\triangleright$ Knowledge Distillation Phase}}
       \STATE Each client trains $\theta_{n}$ on $\{x_k \mid k \in S_t\}$ using $\{Z_k\}$ as soft labels
   \ENDFOR
\end{algorithmic}
\end{algorithm}

In this paper, we primarily focus on three PDA-FD frameworks: \textbf{\textit{FedMD}}~\cite{li2019fedmd}, \textbf{\textit{DS-FL}}~\cite{itahara2021distillation} and\textbf{\textit{ Cronus}}~\cite{chang2019cronus}. 
Each has a specific customization of the procedure demonstrated in Algorithm \ref{alg:FD_algorithm}.
In FedMD, each client $n$ needs to train their local model $\theta_{n}$ on the public dataset $D_{pub}$ until convergence before the collaborative training.
In DS-FL, the server employs the entropy reduction aggregation (ERA)\cite{itahara2021distillation} algorithm as $f_{agg}$ during the communication phase. 
ERA accelerates convergence and enhances the robustness of DS-FL in non-IID data distribution scenarios.
In Cronus, the server utilizes the mean estimation algorithm proposed by Diakonikolas \etal~\cite{diakonikolas2017being} for logits aggregation algorithm $f_{agg}$ to enhance robustness.

\subsection{Privacy Risks Analysis}
Federated Distillation~\cite{chang2019cronus, li2019fedmd, itahara2021distillation} relies on clients and server using inference logits from public datasets to facilitate knowledge transfer and improve model performance. 
However, we find that in the existing PDA-FD algorithms~\cite{chang2019cronus, li2019fedmd, itahara2021distillation}, as shown in Algorithm~\ref{alg:FD_algorithm}, to better transfer knowledge from each client's private model, it is often necessary to train the private model on the private dataset for few epochs during local updates phase before communication phase in collaborative training rounds. 
We believe that this behavior increases the private model's overfitting level to its own private dataset, which inherently raises privacy leakage risks~\cite{yeom2018privacy}. 

\begin{figure}[h]
\centering
\includegraphics[width=1\linewidth]{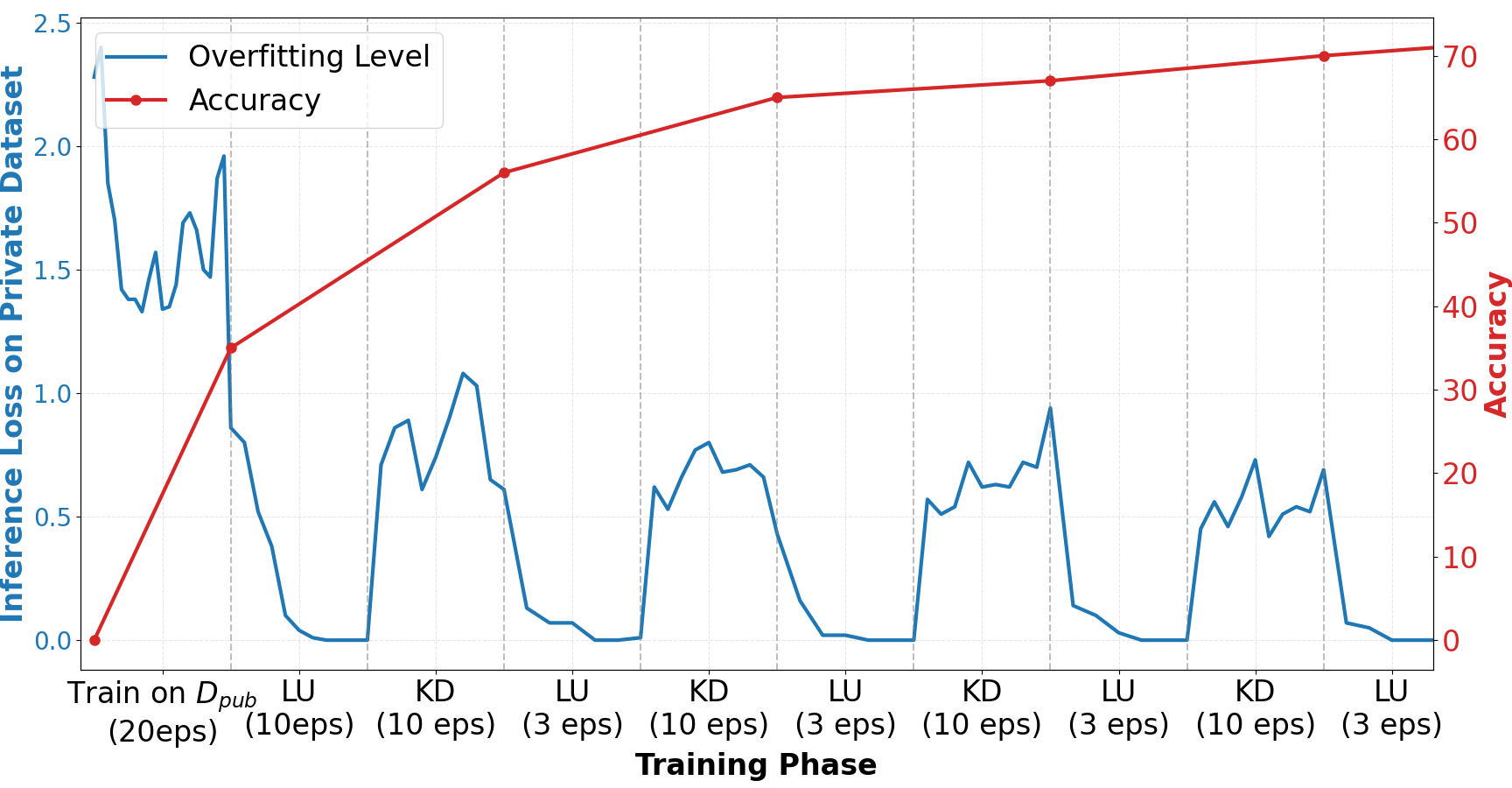}
\caption{Visualizing the client's private model performance trend and overfitting trend to private dataset(members) through collaborative training rounds in FedMD~\cite{li2019fedmd}.}
\label{fig:motivation}
\end{figure}

Our analysis of FedMD training on CIFAR10~\cite{krizhevsky2009learning}, as illustrated in Figure~\ref{fig:motivation}, confirms these concerns by tracking one client's model accuracy and its overfitting to private dataset (measured by inference loss on private dataset) throughout collaborative training. 
Despite the initial training on both public and private datasets, model accuracy continues to improve significantly during collaborative rounds, demonstrating effective knowledge transfer via public datasets. 
However, during each round's local updates(LU) phase, the overfitting level to private data substantially increases. 
More critically, during the subsequent communication phase, which occurs prior to the knowledge distillation(KD) phase, the server needs to interact with client private models through the public dataset to help process and distribute knowledge among all clients. This interaction, combined with the elevated overfitting level, makes privacy attacks feasible. 
Meanwhile, to ensure effective PDA-FD training and facilitate the processing of clients' generated logits, PDA-FD protocols~\cite{li2019fedmd,chang2019cronus,itahara2021distillation} typically require the server to prepare a public dataset for all the clients before training and to select subsets for knowledge distillation in each round.
Therefore, an honest-but-curious server is capable of launching privacy attacks against a target client's private dataset by carefully crafting the public dataset and leveraging the inference logits generated by the client.
The privacy threat posed by such a server to clients should be given significant attention~\cite{gu2023ldia, wang2024graddiff, nasr2019comprehensive}.

\subsection{Training Data Privacy Attacks}
Machine learning models face privacy attacks targeting training data. We focus on two key threats: Label Distribution Inference Attacks (LDIA)~\cite{gu2023ldia, wainakh2021user} and Membership Inference Attacks (MIA)~\cite{shokri2017membership, salem2020updates, carlini2022membership, liu2022membership, ye2022enhanced}, representing privacy violations on the dataset and the sample levels, respectively.
They are also among the most investigated privacy attacks in the field. 
LDIA infers the proportion of different labels in a client's private dataset. For a local model $\theta_{n}$ trained on dataset $D_{n} = \bigcup_{m=1}^{M} D_n^m$, where $D_n^m$ represents data of label $m$, the ground truth label distribution is:
$    \mathbf{p} = \left( p_1, p_2, \dots, p_M \right), \quad p_m = \frac{|D_n^m|}{|D_n|}$.
The attack function maps from the observable information to an estimated label distribution:
    $\mathcal{A} : \theta_n \mapsto \mathbf{\hat{p}} = \left( \hat{p}_1, \hat{p}_2, \dots, \hat{p}_M \right)$.

In MIA, the attacker determines whether a data sample $(x,y)$ belongs to the target model $\theta$'s training dataset. This attack is defined as:
    $\mathcal{A} : x, \theta \mapsto \{ 0, 1\}$,
where 1 indicates member. MIA exploits the behavioral differences of models on training(member) versus non-training data(non-member), leveraging higher confidence or lower loss on training samples due to overfitting~\cite{yeom2018privacy, shokri2017membership}.

\section{Methodology}
\label{sec:method}

\subsection{Threat Model}

\BfPara{Adversary Knowledge} We investigate PDA-FD in the context of Horizontal FL (HFL), where clients' private data's label distributions can be non-IID.
We consider a threat model where the server acts as an \textit{honest-but-curious}~\cite{paverd2014modelling} adversary.
The server attacker is not allowed to modify the learning process or affect the performance of FD training but gets to select public dataset members used for knowledge transfer; this privilege for the server is common in PDA-FD frameworks~\cite{li2019fedmd, itahara2021distillation, chang2019cronus}. The server can select members of the public dataset in each collaborative training round.
In each training round, the server can only black-box access to clients' models, and it only interacts with client models by running inferences on public data. 
The server can only obtain logit vectors or prediction vectors of every public data sample each client model provides without visibility into the private models' weights or architectures.

\BfPara{Adversary's Objective} The server aims to infer sensitive information about a target client's private dataset, specifically label distribution information and membership information, through two main attack vectors: LDIA and MIA.

\subsection{Label Distribution Inference Attack}
\label{sec:ldia_method}
In Federated Distillation, the server only has black-box access by using a public dataset in PDA-FD.
The core idea behind our proposed LDIA is that the logits or prediction values produced by a client's model still carry information about the distribution of labels in its training data.
This is due to the tendency of neural networks to overfit their training distribution, even when regularization techniques are applied~\cite{yeom2018privacy}.

To validate this idea, we conduct a motivating experiment using the CIFAR-10 dataset~\cite{krizhevsky2009learning}.
We sample a subset of data with a non-IID label distribution from CIFAR-10 as the training dataset $D_{train}$. 
Subsequently, we train a deep neural network model $\theta$ on this training set.
We use the trained model $\theta$ to infer on data $x$ from the CIFAR-10 test dataset, which consists of ten different labels, to obtain the logits vector $z_{\theta}(x)$ from these inference results. 
We also apply the softmax function to get the posterior probabilities vector $v_x$ of each image prediction:
\begin{equation}
    v_{x,i} = \frac{e^{z_\theta(x)_i}}{\sum_{j=1}^{10} e^{z_\theta(x)_j}}
\label{eq:softmax_function}
\end{equation}
, where $v_{x,i}$ is the probability for the $i-$th label.
For each label, we have 500 data samples for model inference and calculate the mean posterior probability vector across all the samples
$V_{mean} = \frac{1}{|S|} \sum_{k \in S} v_{x_k}$
, where $S$ represents the subset for each label.

\begin{figure}[ht]
    \centering
    \includegraphics[width=0.9\linewidth]{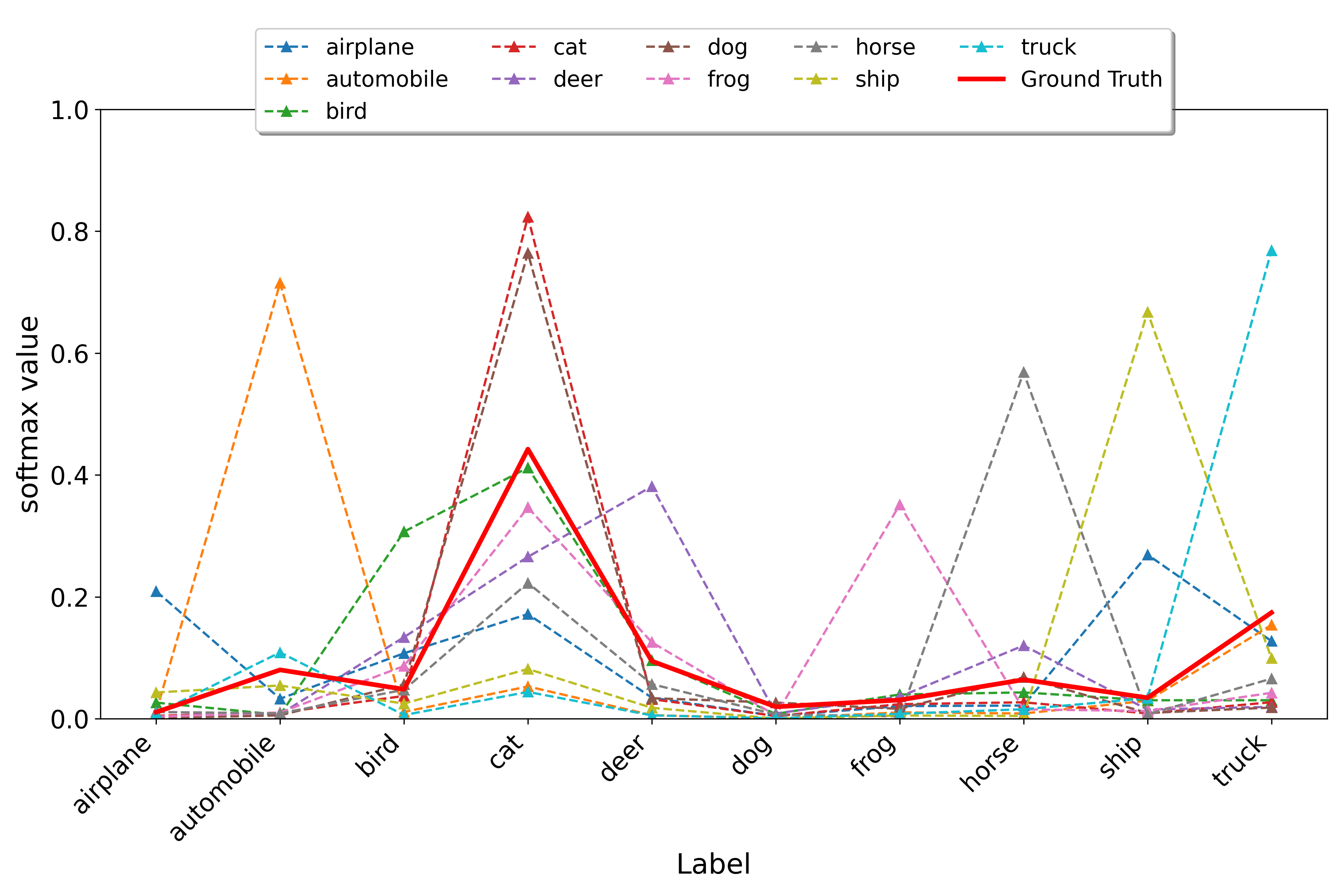}
    \caption{The mean vector of posterior probability vectors $V_{mean}$ predicted by the model $\theta$ on 500 data samples with the same label in CIFAR10.}
    \label{fig:posterior_probabilities_vector}
\end{figure}
Figure \ref{fig:posterior_probabilities_vector} illustrates the mean posterior probability vector $V_{mean}$ for predictions made by model $\theta$ on the subset of data for each label, alongside the label distribution of the training dataset $D_{train}$ used for model $\theta$.
As seen in the figure, the labels ``cat'' and ``truck'' have a larger proportion among all labels in the training dataset $D_{train}$.
However, they constantly appear with higher probabilities in the model's predictions, even for samples of other classes. 
For example, when the model is presented with ``horse'' images, the average prediction probabilities are $0.57$, $0.22$, and $0.07$ for classes ``horse'', ``cat'' and ``truck'', respectively, which are the top-3 predictions.
In this case, the model assigns higher probabilities to the over-represented classes, i.e., ``cat'' and ``truck'', over the other incorrect classes.
This indicates that the model is able to retain the distribution of its training dataset to some extent. 
Such an observation suggests that in FD, where clients share logits or probability values of their models on public data samples, a malicious server could potentially infer the label distribution of clients' private training data.
Even though the public data may have a different distribution, the clients' models will still exhibit biases reflective of their training data distributions.

\begin{figure}
    \centering
    \includegraphics[width=0.8\linewidth]{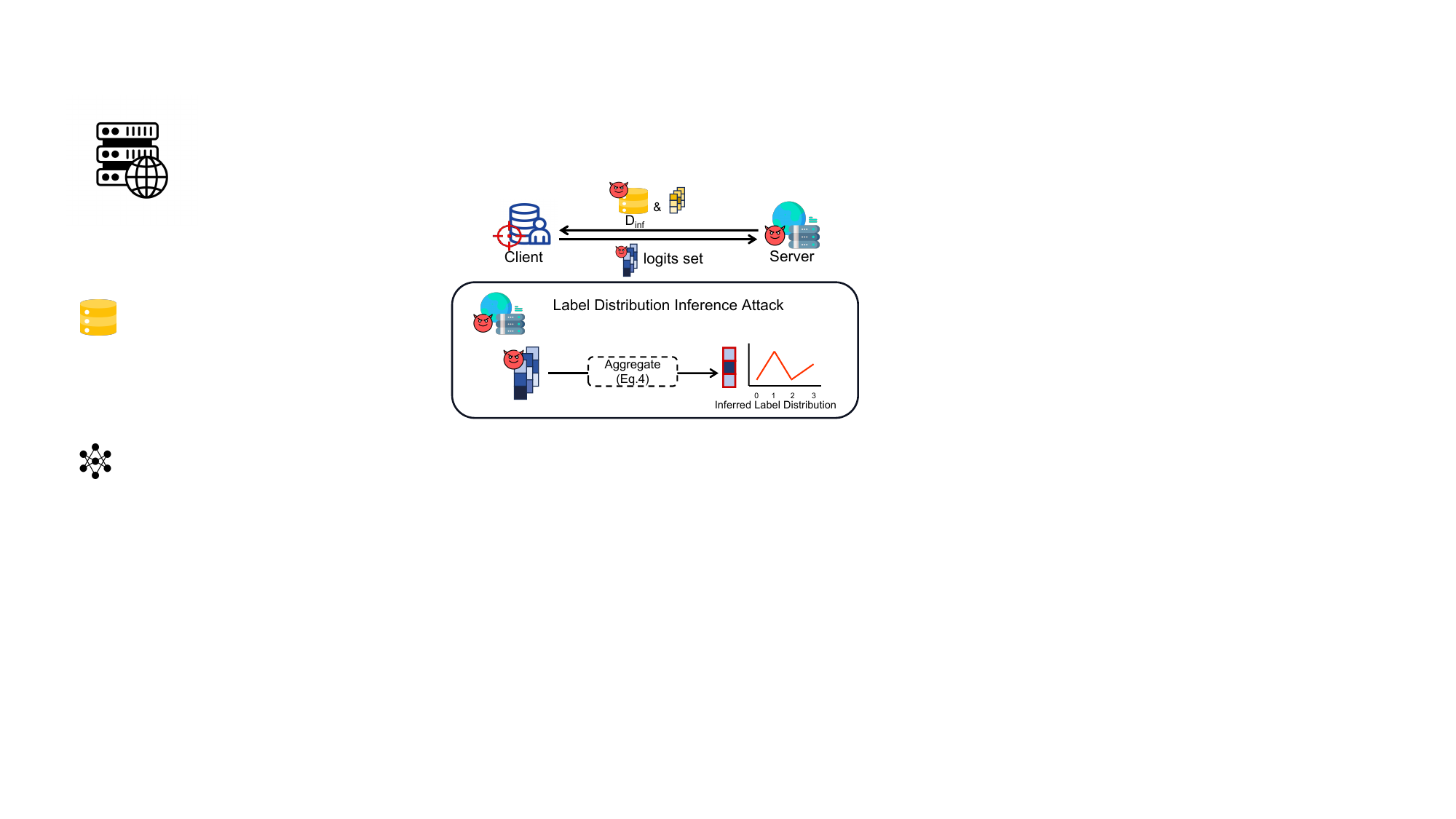}
    \caption{Workflow of Label Distribution Inference Attack.}
    \label{fig:LDIA_workflow}
\end{figure}

With all these positive verification experiment results, we design a LDIA method in PDA-FD that consists of the following steps:
(1) As shown in Figure~\ref{fig:LDIA_workflow}, the server selects a subset of the public dataset $D_{pub}$ for each round of FD, which is denoted as $D_{inf}$.
This selection needs to ensure that the samples are evenly distributed across different classes, which not only helps minimize bias in LDIA but also enhances FD's performance. 
(2) During the communication phase, each client performs inference on the selected public data samples $\{x_k \mid k \in S_t\}$.
The generated logits are sent to the server.
(3) The server receives logits from all the participating clients and selectively aggregate the logits sent by the target client.
The server uses the resulting vector to infer the label distribution of the target client's private dataset.

Formally, the LDIA process can be expressed as:
\begin{equation}
    \hat{p} = \frac{1}{\mid D_{inf} \mid} \sum_{x \in D_{inf}} softmax(z_{\theta}(x))
\label{eq:ldia}
\end{equation}
, where $\hat{p}$ denotes the inferred label distribution, $\theta$ is the target model, and $z_{\theta}(x)$ represent the logits vector obtained by model $\theta$ when inferring on data $x$.
To enhance accuracy and robustness, we average the inferred label distributions over $N$ rounds and report the averaged distribution as the final LDIA result. This mitigates the impact of potential anomalies or fluctuations in individual rounds, thus improving overall stability.

\subsection{Membership Inference Attacks}
MIA in machine learning aims to determine whether a specific data sample was used to train a model.
Recent works have demonstrated the feasibility of MIA in traditional FL and centralized machine learning setting, where adversaries have either white-box access to the model's parameters and gradients or access to shadow datasets matching the target model's training dataset's data distribution for performing MIA~\cite{carlini2022membership, watson2021importance, shokri2017membership}.
However, in the context of PDA-FD, the server is limited to black-box interactions with the target client's model, and lacks knowledge of clients' private data distributions, making it challenging to obtain appropriate shadow datasets.
To address this challenge, we propose adapting and enhancing existing MIA approaches by combining them with the unique characteristics of the FD process.

Traditional MIA techniques often exploit the observation that machine learning models behave differently on data they are trained on versus data they aren't.
The difference typically manifests as higher confidence or lower loss on training samples.
In our proposed attack, we extend the state-of-the-art MIA technique, Likelihood Ratio Attack (LiRA), proposed by Carlini \etal~\cite{carlini2022membership}.

Offline LiRA uses multiple reference ("out") models to establish a baseline prediction distribution for membership inference through hypothesis testing. 
The key assumption is that member samples have statistically different prediction scores between target and reference models, while non-members exhibit similar predictions across all models.
For a target sample $(x,y)$, LiRA first queries all reference models to obtain posterior probabilities, which are then standardized using the following scaling function:
\begin{equation}
    \phi(f_\theta(x)_{y}) = \log\left(\frac{f_\theta(x)_{y}}{1 - f_\theta(x)_{y}}\right)
    \label{eq:lira_rescale}
\end{equation}
where $f_{\theta}(x)_y$ denotes the posterior probability from a reference model $\theta$.
The scaled scores from reference models are fitted to a Gaussian distribution $\mathcal{N}(\mu_{\text{out}}, \sigma_{\text{out}}^2)$. The membership probability $\lambda$ is then computed by comparing the target model's scaled score $\phi(f_{\theta_t}(x)_y)$ with this distribution:
\begin{equation}
    \lambda = 1 - \Pr[Z > \phi(f_{\theta_t}(x)_y)], \text{where } Z \sim \mathcal{N}(\mu_{\text{out}}, \sigma_{\text{out}}^2).
    \label{eq:lira_result}
\end{equation}
A higher scaled score $\phi(f_{\theta_t}(x)_y)$ relative to $\mu_{\text{out}}$ indicates a higher probability of $(x,y)$ being a member of the training dataset.

\begin{figure}[ht]
\centering
  \begin{subfigure}[m]{0.49\linewidth}
  \centering
    \includegraphics[width=\linewidth]{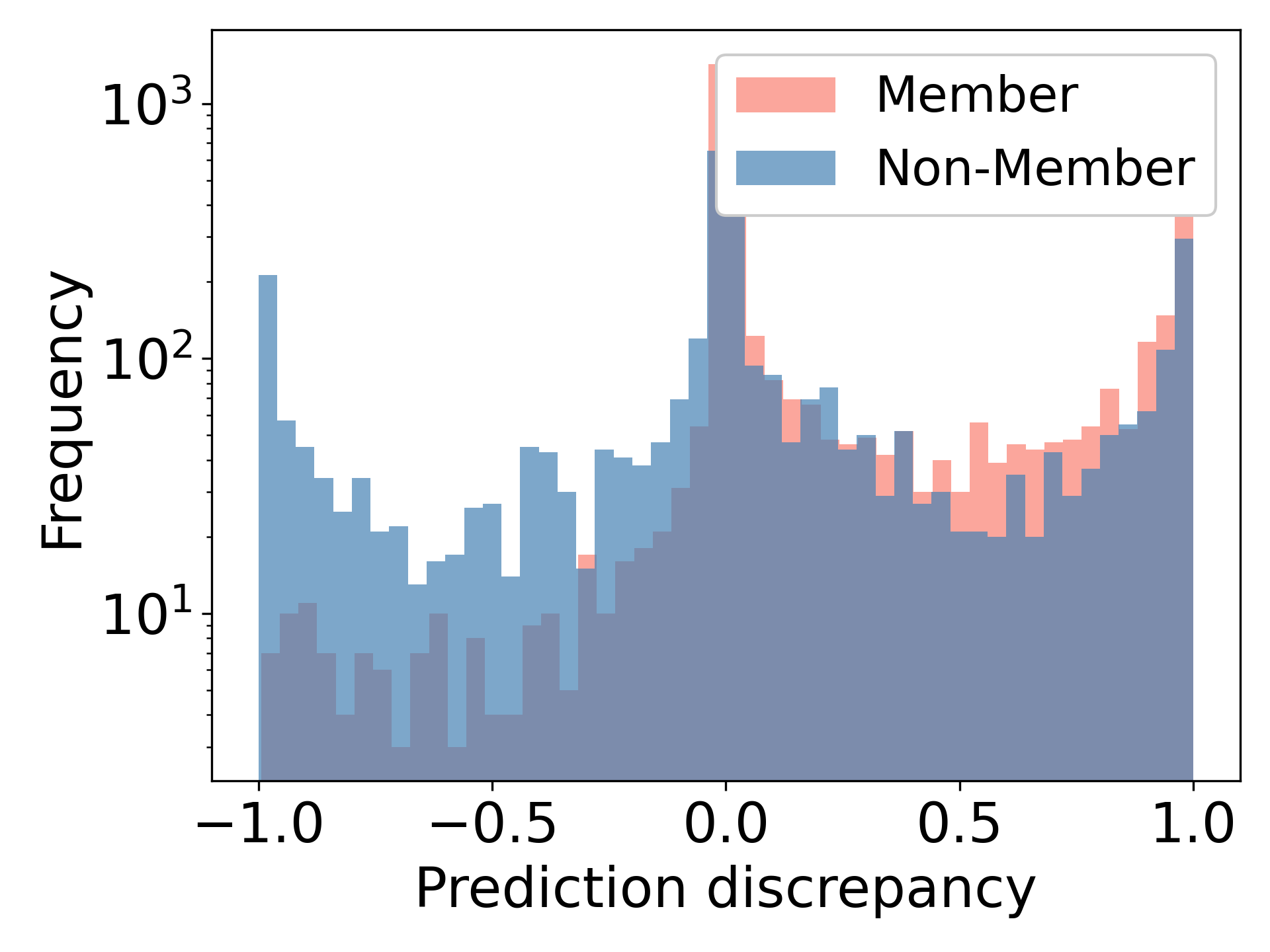}
    \caption{Reference model trained with $D_{ref}$}
    \label{fig:regular_reference_model}
  \end{subfigure}
  \hfill
  \begin{subfigure}[m]{0.49\linewidth}
  \centering
    \includegraphics[width=\linewidth]{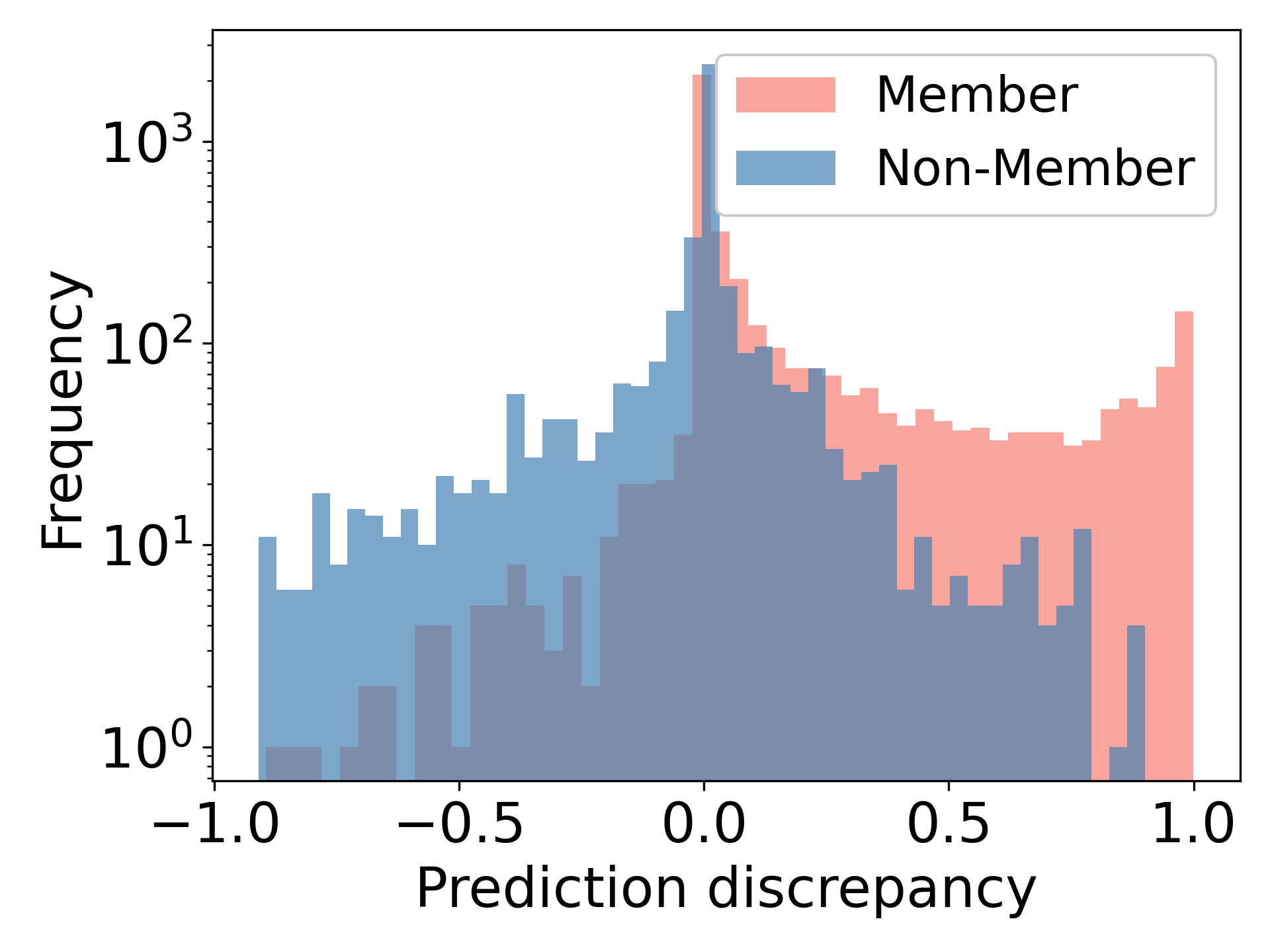}
    \caption{Reference model distilled by using $D_{ref}$}
    \label{fig:distilled_reference_model}
  \end{subfigure}
  \caption{Prediction discrepancies between target and reference models across members and non-members on CIFAR 10. The target model is trained on dataset $D_{target}$ while the adversary can only obtain $D_{ref}$, where $D_{target}$ and $D_{ref}$ have different distributions.}
  \label{fig:distlled_reference_model_comparison}
\end{figure}

While LiRA is a powerful and effective technique for MIAs, directly applying it in the context of PDA-FD presents challenges.
In traditional LiRA, the attacker needs the ability to train multiple reference models that mimic the target model's behavior but behave differently on the target samples, specifically, prediction discrepancies between target and reference models should be larger for members than for non-members.
However, training such models requires access to a shadow dataset with a distribution similar to the target model's training data~\cite{watson2021importance, carlini2022membership}.
As shown in Figure \ref{fig:regular_reference_model}, when the data distribution of the reference model's training dataset differs from that of the target model's training dataset, the attacker cannot distinguish members based on prediction discrepancies between reference and target models for members versus non-members. The prediction discrepancy is quantified as the difference between the target model's prediction probability and the mean of reference models' prediction probabilities for the target sample.

\begin{figure}
    \centering
    \includegraphics[width=1\linewidth]{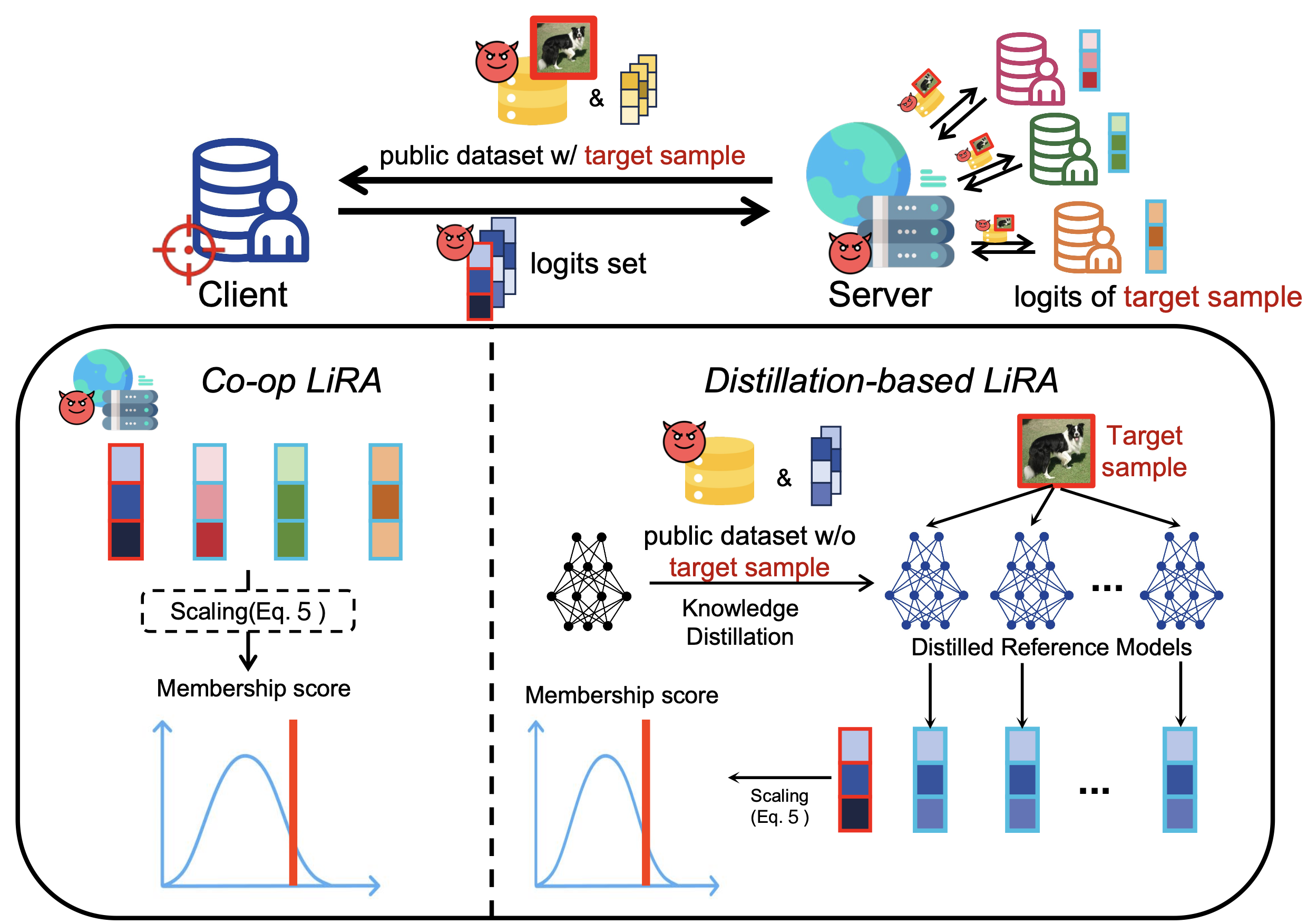}
    \caption{Workflows of two proposed Membership Inference Attacks(Co-op LiRA and Distillation-based LiRA).}
    \label{fig:MIA_workflow}
\end{figure}

In the PDA-FD context, the server faces significant challenges in obtaining such a shadow dataset:
(1) The FD framework allows clients to have non-IID private datasets in the label space.
Without knowledge of the specific label distributions in each client's private dataset, the server cannot accurately sample shadow datasets that match the characteristics of the target client's data.
(2) In FD, the public dataset can be unlabeled, particularly in semi-unsupervised learning scenarios.
In that case, the server simply cannot use these datasets without labels for training shadow models, as label information is essential for mimicking the target model's behaviors.
To address these unique challenges, as shown in Figure \ref{fig:MIA_workflow}, we design and implement two variants of offline LiRA: Co-op LiRA and Distillation-based LiRA.
Section~\ref{sec:indirect_attacks} also briefly explores how these approaches can be combined with indirect attack techniques~\cite{long2020pragmatic, wen2022canary}, further demonstrating the robustness of our proposed attacks.

\subsubsection{\textbf{Co-op LiRA}}
Our proposed Co-op LiRA introduces a novel approach that leverages our LDIA method as a prerequisite to enable effective membership inference attacks. Unlike traditional MIA techniques that require shadow datasets with similar distributions to the target, our method uniquely exploits the collaborative nature of federated distillation environments.
The key innovation of Co-op LiRA lies in its ability to identify suitable reference models without additional training. 
We aim for these reference models to have training datasets with data distributions closer to the target model's distribution, thereby better mimicking the target client model's behavior~\cite{shokri2017membership}.
First, the server conducts LDIA (as described in Section~\ref{sec:ldia_method}) to determine label distributions across all clients.
In the HFL setting, clients with similar label distributions likely possess 
closer overall data distributions.
This is because the data distribution may differ in two aspects: the feature and label space. 
In the definition of HFL, clients' data share the same feature space. In this context, more similar label distributions imply more similar overall data distributions.
These clients' models can then serve as effective reference models for performing LiRA against the target client~\cite{shokri2017membership, carlini2022membership}.
While we notice that when target samples appear in other clients' private datasets, those clients' models cannot serve as reference models. 
However, this limitation is minimized in real-world FD deployments where clients typically have mostly disjoint training datasets to benefit from collaborative learning.

Algorithm \ref{alg:MIA_algorithm} outlines the co-op LiRA process in PDA-FD:
(1) When the server prepares the public dataset, it incorporates a set of potential target samples into the public dataset. 
During the communication phase, the server selects subsets containing these target samples for knowledge transfer, allowing it to obtain posterior probabilities of the subset of public dataset from all clients' models, including the target client.
(2) The server conducts LDIA on all clients.
(3) The server calculates the KL-divergence between the label distributions of the target client and each remaining client, selecting clients as reference models when their KL-divergence falls below threshold $\beta$ (0.1).
(4) The server performs LiRA hypothesis test by comparing reference models' and target client's posterior probabilities on the target sample to determine target sample's membership probability.

Co-op LiRA eliminates the need for the attacker to train multiple reference models. 
However, the drawback of the method is that it becomes ineffective when very few clients have training data of the same or similar distribution to the target client.

\begin{algorithm}
\caption{Co-op LiRA and Distillation-based LiRA. The shadow model preparation phase is implemented in \textbf{lines 6-12 for Co-op LiRA} and \textbf{lines 14-18 for Distillation-based LiRA}.}
\label{alg:MIA_algorithm}
\begin{algorithmic}[1]
    \REQUIRE public dataset $D_{pub}$, target sample $(x,y)$, target client private model $\theta_{target}$, clients private models $\{\theta_n\}_{n=1}^N$, scaling function $\phi$, KL-divergence threshold $\beta$, number of reference models $K$.
    \STATE $\mathcal{M}_{ref} \leftarrow \{\}$, $Conf \leftarrow \{\}$
    \STATE {\color{blue}\textit{$\triangleright$ Communication phase in FD}}
    \STATE $D_{pub} \leftarrow D_{pub} \cup \{(x,y)\}$ 
    \STATE Server send $D_{pub}$ to all clients and receive $\{f_{\theta_n}(D_{pub})\}_{n=1}^N$
    \STATE {\color{blue}\textit{$\triangleright$ Co-op LiRA}}
    \STATE $\hat{p}_{target} \leftarrow \text{LDIA}(\theta_{target})$ 
    \FOR{$\theta_n \in \{\theta_n\}_{n=1}^N \setminus \{\theta_{target}\}$}
        \STATE $\hat{p}_n \leftarrow \text{LDIA}(\theta_n)$
        \IF{$\text{KL}(\hat{p}_{target}, \hat{p}_n) < \beta$}
            \STATE $\mathcal{M}_{ref} \leftarrow \mathcal{M}_{ref} \cup \{\theta_n\}$
        \ENDIF
    \ENDFOR
    \STATE {\color{blue}\textit{$\triangleright$ Distillation-based LiRA}}
    \STATE $D_{distill} \gets \text{Random sample}(D_{pub} \setminus \{(x,y)\})$ 
    \FOR{$k=1$ to $K$}
        \STATE $\theta_k \leftarrow \text{Distill} (\theta_{target}, D_{distill})$
        \STATE $\mathcal{M}_{ref} \leftarrow \mathcal{M}_{ref} \cup \{\theta_k\}$
    \ENDFOR
    \FOR{$\theta \in \mathcal{M}_{ref}$}
        \STATE $Conf \leftarrow Conf \cup \{\phi(f_{\theta}(x)_y)\}$
    \ENDFOR
    \STATE $\mu_{out} \leftarrow \text{mean}(Conf)$, $\sigma^{2}_{out} \leftarrow \text{var}(Conf)$
    \RETURN $\lambda \leftarrow 1 - \Pr[Z > \phi(f_{\theta_{target}}(x)_y)]$, where $Z \sim \mathcal{N}(\mu_{out}, \sigma^{2}_{out})$
\end{algorithmic}
\end{algorithm}

\subsubsection{\textbf{Distillation-based LiRA}} To address the limitation of co-op LiRA, we proposed another extension of LiRA called distillation-based LiRA.
It is challenging for the server to obtain a shadow dataset which has same data distribution with the target's model's training dataset to train a reference model; we instead choose to use knowledge distillation to generate a student model, which is then used as a reference model for the MIA. 
On the one hand, the generated student model should behave similarly to the target model, exhibiting close prediction scores on non-member samples.
On the other hand, it produces different prediction scores on member data that differ from those from the teacher model, since the student model is trained through knowledge distillation without direct exposure to the training dataset, whereas the teacher model, trained directly on member samples, exhibits a degree of overfitting to these samples.
Previous research on MIA~\cite{jagielski2024students} has demonstrated that the student model generated through knowledge distillation can potentially encode membership information of the teacher model. 
However, we find significant prediction discrepancies between student and teacher models when performing inference on the teacher model's member samples. As shown in Figure \ref{fig:distilled_reference_model}, the prediction discrepancies between the teacher and the student model are significant for some members while remaining relatively small for non-members. 
This distinct pattern enables these distilled student models to serve as effective reference models for offline LiRA. 
These reference models allow attackers to identify members with high prediction discrepancies while maintaining a low False Positive Rate, thus achieving high True Positive Rate (TPR) at low False Positive Rate (FPR), which serves as a critical performance metric for successful membership inference attacks\cite{carlini2022membership, watson2021importance}. 
We validate this characteristic of knowledge distillation-based reference models in Section \ref{sec:MIA_eval}.

Algorithm \ref{alg:MIA_algorithm} outlines the distillation-based LiRA process in PDA-FD:
(1)  When the server prepares the public dataset, it incorporates a set of potential target samples into the public dataset.
During the communication phase, the server selects subsets containing the target samples for knowledge transfer.
The server obtains the posterior probabilities predicted by the target client on all the data samples. 
These probability values could be used as soft labels to distill reference models.
(2) To create multiple reference models, the server randomly samples the distillation dataset of distilled model from the public dataset except the target samples. Using the target client as the teacher model, the server distills multiple reference models.
(3) The server then infers membership of the target samples by performing LiRA using the created reference models.
This approach is more robust to heterogeneous data distributions among clients but comes at the cost of additional computational overhead for the distillation process.

\section{Evaluations}
\label{sec:eval}
In this section, we conduct a series of experiments to evaluate the privacy leakage in PDA-FD by conducting the proposed LDIA and MIA.
Our experiments span multiple datasets, various PDA-FD frameworks~\cite{chang2019cronus, li2019fedmd, itahara2021distillation}, and different usage scenarios. 
Our experiments demonstrate four key aspects: (1) the overall effectiveness of our proposed LDIA and MIA; (2) the impact of varying non-IID data distributions; (3) the impact of different PDA-FD frameworks; and (4) the effect of the number of collaborative training rounds.

\subsection{Experiment Setup}
\subsubsection{Datasets}
In our experiments, we utilize the following image datasets that are commonly used to test the performance of different FD frameworks: CIFAR-10\cite{krizhevsky2009learning}, CINIC-10\cite{darlow2018cinic}, Fashion-MNIST\cite{xiao2017fashion}.
Additionally, for the completeness of the experiments, we also use a tabular dataset:Purchase\cite{acquire-valued-shoppers-challenge}.
In our experiments, we partition each dataset into a 4:1 ratio for clients' training sets $D_{train}$ and the public dataset $D_{pub}$.
To align with the previous FD frameworks~\cite{yang2022fd, li2019fedmd}, we also configure a scenario where there is a data distribution discrepancy between the public dataset and the clients' private datasets.
In this scenario, we use CIFAR-10 for client training and CIFAR-100 as the public dataset to simulate distribution shifts.
The details of the specific partitioning of datasets and the number of classes used in our experiments can be found in Appendix~\ref{sec:datasets_division}.
In our MIA experiments, to ensure adequate private data for each client, we select an equal number of samples from the test dataset to serve as non-members. 

In our experiments, we create 10 clients that participate in the collaborative training. 
To partition the training dataset $D_{train}$ into private datasets $D_n$ for 10 clients, we use Dirichlet distribution $Dir(\alpha)$ with $\alpha$ values of $0.1$, $1$ and $10$ to generate non-IID data distribution across all the clients. 
The smaller the value of $\alpha$, the more imbalanced the label distribution of $D_n$ is. 

\subsubsection{Models.}
For different datasets, the clients in PDA-FD use different model architectures. 
When the private dataset is CIFAR-10, the clients train the ResNet-18 models~\cite{he2016deep}. 
For CINIC-10, the clients train the MobileNetV2 models~\cite{sandler2018mobilenetv2}.
For Fashion-MNIST datasets, the clients' local models employ a CNN architecture with four convolutional layers. 
When training with the Purchase dataset, the clients train MLP models consisting of three fully connected layers.
As the heterogeneity in client model architectures does not affect our attack methodology\cite{carlini2022membership}, we adopted identical model structures across all clients.

\subsubsection{LDIA Metrics.}
To evaluate the effectiveness of the proposed LDIA, we adopt the same metrics employed in previous LDIA research~\cite{gu2023ldia, salem2020updates}. 
In the equations for calculating these metrics, $\hat{p}$ denotes the inferred label distribution, $p$ denotes the ground truth label distribution, $m$ represents a specific label, and $M$ denotes the number of labels:
\begin{itemize} [leftmargin=*]
    \item \textbf{Kullback-Leibler divergence.}
    Kullback-Leibler divergence(KL divergence) between the ground truth label distribution and the inferred label distribution can be calculated using the following equation: 
    \begin{equation}
        Dis_{KL-div}(\hat{p}, p) = \sum_{m=1}^M \hat{p}_m log(\frac{\hat{p}_m}{p_m})
    \label{eq:KL_metrics}
    \end{equation}
    This metric represents the similarity between the inferred label distribution and the ground truth label distribution. 
    A smaller KL divergence indicates that the two distributions are more closely aligned.
    \item \textbf{Chebyshev distance.}
    The Chebyshev distance represents the maximum error between the inferred label distribution and the ground truth label distribution for each target client in an LDIA:
    \begin{equation}
        Dis_{Cheb}(\hat{p}, p) = max_m\mid \hat{p}_m - p_m \mid
    \label{eq:Cheb_metrics}
    \end{equation}
    A smaller Chebyshev distance indicates a higher reliability of the LDIA results.

\end{itemize}

\subsubsection{MIA Metrics.}
Same as previous efforts on MIA~\cite{carlini2022membership,shokri2017membership,watson2021importance} , we employ the following metrics to evaluate the effectiveness of the proposed MIAs:
\begin{itemize} [leftmargin=*]
    \item \textbf{TPR at low FPR.} Carlini \etal~\cite{carlini2022membership} suggested using TPR at low FPR to measure MIA. 
    A higher TPR in the low FPR region indicates greater precision of the MIA, which also implies that the attack is more reliable.
    \item \textbf{Balance accuracy and AUC.} These two metrics assess the overall performance of MIA. 
    Balanced accuracy measures the attacker's ability to correctly predict true positives and true negatives across all members and non-members. 
    AUC quantifies the area beneath the ROC curve of the MIA results. 
    It offers a comprehensive measure of the attack's discriminative power across various classification thresholds.
\end{itemize}

\subsubsection{PDA-FD Frameworks.}
To comprehensively evaluate the privacy leakage in the PDA-FD setting, we evaluate three different PDA-FD frameworks in our experiments: FedMD~\cite{li2019fedmd}, DS-FL~\cite{itahara2021distillation}, and Cronus~\cite{chang2019cronus}. 
Each of these FD frameworks employs a distinct approach to enhance the robustness of the FD algorithms.
As mentioned in Section~\ref{sec:background_fd}, they behave differently during the local updates phase and use different aggregation algorithms during the communication phase.
In order to optimize the performance of all the PDA-FD frameworks, we should carefully select the number of training epochs in each round.
Following the approach suggested by Li \etal in FedMD~\cite{li2019fedmd}, we initially train the local models to convergence on the public and private datasets before transitioning to the shorter update cycles during the distillation phase.
Specifically, in the first round of local updates, each client performs 20 epochs of training. 
In the subsequent rounds, this is reduced to 5 epochs of training for each client. 
The knowledge distillation phase consists of 10 epochs of training for each client.
To reduce communication cost, we randomly select 5000 data samples from the public dataset during each communication phase for the CIFAR-10, CIFAR10/CIFAR100, Fashion-MNIST and Purchase datasets, aligned with previous PDA-FD studies~\cite{li2019fedmd}.
The CINIC-10 dataset, given its difficulty level, uses a set of 10000 randomly selected samples to ensure adequate knowledge transfer.
Table \ref{tab:FD_performance} presents the performance of the three PDA-FD frameworks across various Dirichlet distributions and five distinct dataset settings.
\begin{table}[ht]
\caption{Performance of the PDA-FD Frameworks.}
\label{tab:FD_performance}
\centering
\scriptsize
\resizebox{1\linewidth}{!}{%
    \begin{tabular}{lcccccc}
    \toprule
    \multirow{2}{*}{Datasets} & 
    \multirow{2}{*}{Setting} & 
    \multirow{2}{*}{Local accuracy} & 
    \multicolumn{3}{c}{Federated accuracy} \\ 
    \cmidrule(lr){4-6}
     & & &
     FedMD & DS-FL & Cronus \\
     \midrule
     CIFAR10 
     & $\alpha$=10  & 54.59\% 
     & 76.61\%  & 71.31\% & 70.81\% \\
     & $\alpha$=1   & 46.06\% 
     & 75.38\% & 68.45\% & 67.90\% \\
     & $\alpha$=0.1  & 22.75\%
     & 60.55\% & 45.01\% & 42.24\% \\
    \midrule
     CIFAR10 
     & $\alpha$=10   & 53.24\% 
     & 72.34\% & 69.54\% & 68.42\% \\
     /CIFAR100 & $\alpha$=1 & 45.49\% & 
     68.41\% & 65.90\% & 64.26\% \\
     & $\alpha$=0.1 & 23.31\%  & 
     49.89\% & 43.55\% & 43.43\% \\
    \midrule
     CINIC10 
     & $\alpha$=10& 39.31\%  
     &67.72\%  &64.21\%  & 62.92\% \\
     & $\alpha$=1& 33.37\%   
     &62.91\%  &57.02\%  & 56.49\% \\
     & $\alpha$=0.1& 20.45\% 
     &41.02\%  &38.48\%  &37.89\%  \\
     \midrule
     Fashion & $\alpha$=10    & 78.80\% 
     & 88.68\% & 87.96\% & 87.62\% \\
     -MNIST  & $\alpha$=1   & 69.35\%
     & 87.98\% & 85.25\% & 84.98\% \\
     & $\alpha$=0.1 & 19.58\% 
     & 80.62\% & 52.45\% & 56.34\% \\
    \midrule
     Purchase 
     & $\alpha$=10  & 82.56\% 
     &  94.58\% & 88.35\% & 88.62\%\\
     & $\alpha$=1 & 72.73\% 
     & 94.01\% & 86.83\% & 89.65\% \\
     & $\alpha$=0.1  & 52.18\% 
     & 91.80\% & 72.55\% & 67.34\% \\
     \bottomrule
    \end{tabular}%
    }
\end{table}

\subsection{Experiment Results of LDIA}
In the LDIA experiments, the PDA-FD server infers the label distribution of all clients' private datasets during the communication phase in each round.
To ensure robustness and account for potential fluctuations, we compute the final LDIA result for each client by averaging the server's inferred label distribution over 10 collaborative training rounds.
The overall effectiveness of the attack is evaluated by averaging these final results across all clients.
To provide a meaningful benchmark for our LDIA method, we establish a baseline comparison, denoted as ``Random'', following the same approach of previous LDIA research~\cite{gu2023ldia, salem2020updates}. 
This baseline employs randomly generated label distributions for each client's private dataset, serving as a lower bound for attack performance.
Note that for a given dataset and Dirichlet distribution parameter, the private dataset of each client remains constant across different PDA-FD frameworks. 
Therefore, within the same dataset and Dirichlet distribution, there is only one set of Random LDIA results.

\begin{figure}[htbp]
    \centering
    \begin{subfigure}[b]{\linewidth}
        \centering
        \includegraphics[width=1\linewidth]{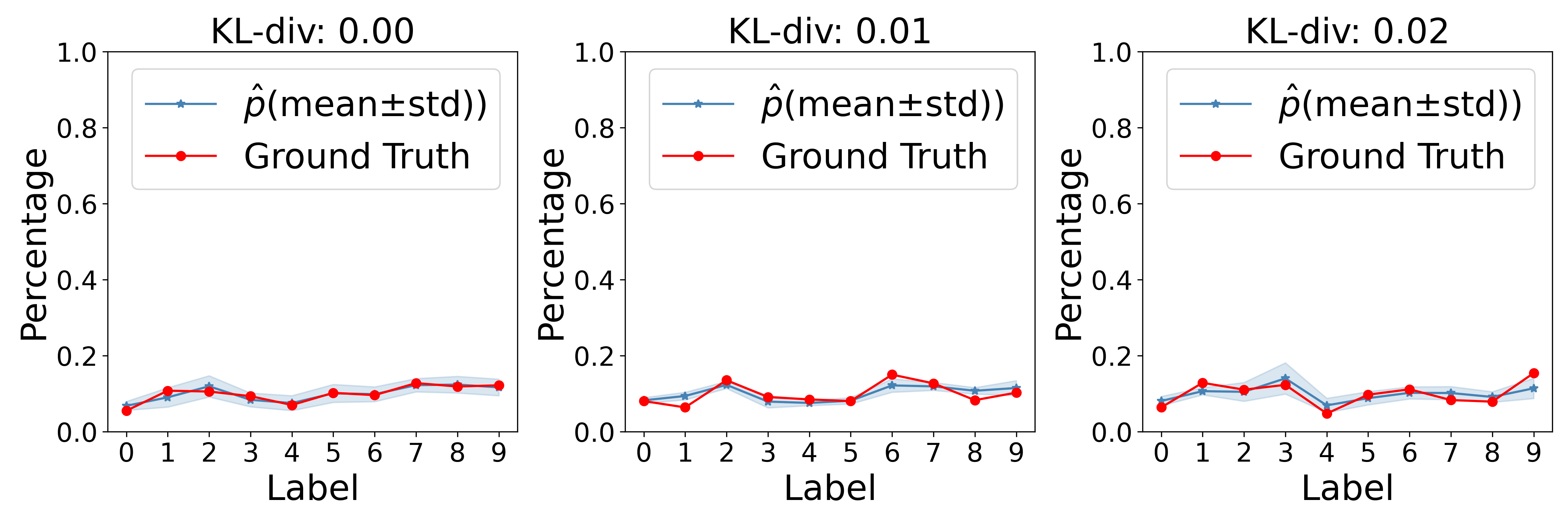}
        \caption{$\alpha=10$}
        \label{fig:subfig1}
    \end{subfigure}

    \begin{subfigure}[b]{\linewidth}
        \centering
        \includegraphics[width=1\linewidth]{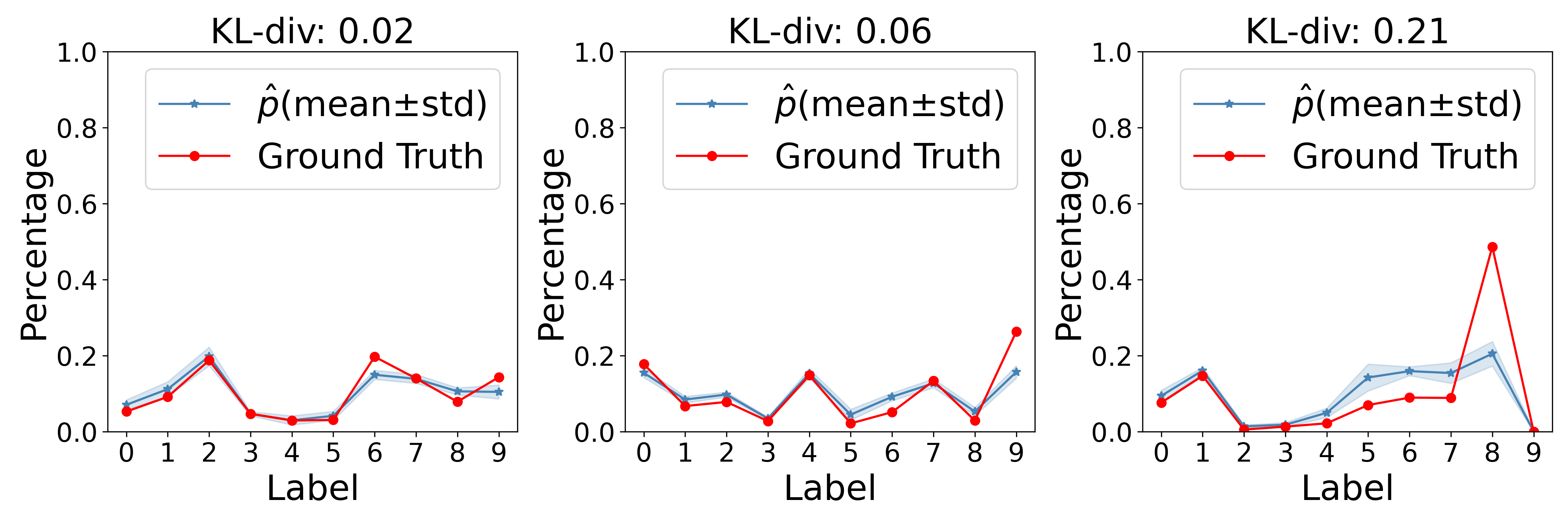}
        \caption{$\alpha=1$}
        \label{fig:subfig2}
    \end{subfigure}
    
    \begin{subfigure}[b]{\linewidth}
        \centering
        \includegraphics[width=1\linewidth]{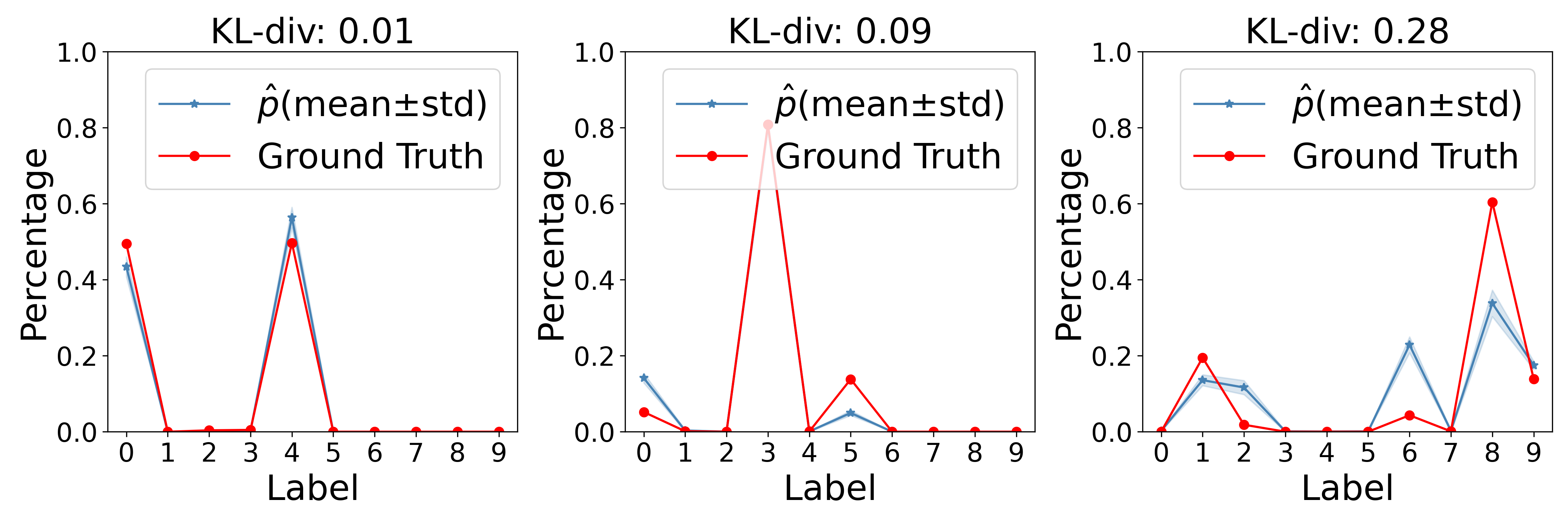}
        \caption{$\alpha=0.1$}
        \label{fig:subfig3}
    \end{subfigure}

    \caption{The LDIA performance of the DS-FL server on the CIFAR-10 dataset, under three distinct Dirichlet distributions. The images depict the best (left), median (center), and worst (right) LDIA results across all client models.}
    \label{fig:ldia_main_result}
\end{figure}

\textbf{Main Result.} Table \ref{tab:ldia_main_result} presents the performance of the proposed LDIA on five different datasets across three PDA-FD frameworks. 
The results demonstrate the server's capability to launch effective LDIA against clients across these datasets, significantly outperforming the random guess baseline on all three key metrics.
For instance, for the DS-FL framework on the CIFAR-10 dataset with $\alpha$=1, our proposed LDIA achieves an average Chebyshev distance of 0.11 and an average KL-divergence of 0.10 across all clients. 
In contrast, the random guess baseline yields substantially higher values: 0.20 and 0.66 for the respective metrics. 
This significant improvement underscores the efficacy of our LDIA in accurately inferring clients' label distributions.
Figure \ref{fig:ldia_main_result} provides a visual representation of the LDIA results for the DS-FL server on the CIFAR-10 dataset, offering a clearer illustration of the experiment results.
We can see from the figure that for the labels whose inferred proportions deviate from the ground truth values, the relative rankings of label frequencies are consistently preserved.
This observation highlights the robustness of the proposed LDIA in capturing the essential structure of label distributions.

\begin{table}[ht]
\caption{Performance of the server in conducting LDIA within the different PDA-FD Frameworks.}
\centering
\setlength{\tabcolsep}{3pt}
\resizebox{1\linewidth}{!}{%
\begin{tabular}{lccccccccc}
\toprule
\multirow{2}{*}{Datasets} & 
\multirow{2}{*}{Setting} & 
\multicolumn{4}{c}{KL divergence} & 
\multicolumn{4}{c}{Chebyshev distance} \\ 
\cmidrule(lr){3-6}
\cmidrule(lr){7-10}
 & & 
 FedMD & DS-FL & Cronus & Random &
 FedMD & DS-FL & Cronus & Random \\ 
 \midrule
 CIFAR10 
 & $\alpha$=10 
 & 0.02 & 0.01 & 0.01 & 0.42
 & 0.04 & 0.03 & 0.02 & 0.13\\
 & $\alpha$=1   
 & 0.17 & 0.10 & 0.08 & 0.66
 & 0.14 & 0.11 & 0.10 & 0.20\\
 & $\alpha$=0.1 
 & 0.15 & 0.11 & 0.07 & 1.93
 & 0.18 & 0.16 & 0.14 & 0.57\\
\midrule
 CIFAR10 
 & $\alpha$=10  
 & 0.07 & 0.05 & 0.06 & 0.40
 & 0.07 & 0.06 & 0.06 & 0.12\\
 /CIFAR100 & $\alpha$=1 
 &0.11 & 0.10 & 0.10 & 0.59
 &0.10 & 0.11 & 0.11 & 0.22\\
 & $\alpha$=0.1 
 &0.11 & 0.10 & 0.08 & 1.59
 &0.15 & 0.13 & 0.14 & 0.51\\
\midrule
 CINIC10
 & $\alpha$=10
 &0.01 & 0.01 & 0.01 &0.64  
 &0.02 & 0.02 & 0.04 &0.15\\
 & $\alpha$=1 
 &0.06 & 0.05 & 0.05 &0.57  
 &0.11 & 0.11 & 0.10 &0.21\\
 & $\alpha$=0.1 
 &0.09 & 0.08 & 0.01 &1.99  
 &0.14 & 0.14 & 0.04 &0.56\\
 \midrule
 Fashion & $\alpha$=10  
 & 0.03 & 0.02 & 0.02 & 0.32
 & 0.04 & 0.04 & 0.04 & 0.12\\
 -MNIST  & $\alpha$=1  
 & 0.14 & 0.12 & 0.12 & 0.55
 & 0.10 & 0.09 & 0.09 & 0.15\\
 & $\alpha$=0.1 
 & 0.21 & 0.05 & 0.06 & 1.54 
 & 0.20 & 0.09 & 0.11 & 0.45\\
\midrule
 Purchase 
 & $\alpha$=10 
 &  0.08 & 0.03 & 0.03 & 0.47
 &  0.06 & 0.05 & 0.05 & 0.13\\
 & $\alpha$=1 
 & 0.27 & 0.14 & 0.15 & 0.68
 & 0.13 & 0.10 & 0.10 & 0.18\\
 & $\alpha$=0.1 
 & 0.64 & 0.14 & 0.14 & 2.11
 & 0.32 & 0.15 & 0.17 & 0.52\\
 \bottomrule
\end{tabular}%
}
\label{tab:ldia_main_result}
\end{table}

\textbf{Different Data Distributions.} 
Our experiments reveal a notable relationship between the effectiveness of the proposed LDIA and the uniformity of clients' label distributions.
Specifically, the LDIA demonstrates lower KL-divergence, Chebyshev distance, and mean $l1$-distance as the clients' label distributions become more uniform.
This trend is clearly illustrated in our experiments using the CIFAR-10 dataset within the DS-FL framework. 
As $\alpha$ increases, indicating a more uniform label distribution across clients, the LDIA achieves better performance.
Conversely, as $\alpha$ decreases, indicating a more skewed distribution, we see an increase in the three key metrics.
Nonetheless, the attack remains effective despite the reduced accuracy.

\textbf{Different PDA-FD Frameworks.} 
Our evaluations also reveal significant differences in the vulnerability of various PDA-FD frameworks to LDIA.
Compared to FedMD, the server can achieve more effective LDIA on clients within the DS-FL and Cronus frameworks.
This can be attributed to the unique training approach employed by FedMD during its first collaborative training round.
In FedMD, clients first train their local models on the public dataset before transitioning to their private dataset.
This process serves as a form of regularization, thus mitigating overfitting to private datasets and, consequently, reducing vulnerability to LDIA.
However, this effect is diminished when private and public datasets differ significantly or when the public dataset is unlabeled.
In these cases, FedMD's LDIA vulnerability becomes comparable to that of the other two PDA-FD frameworks, as evidenced by the results in Table \ref{tab:ldia_main_result} for the CIFAR-10/CIFAR-100 datasets.
The similarity in LDIA vulnerability arises from the data distribution shift between public and private datasets, causing clients' local models to reduce memorization of the public dataset after converging on their private datasets.
As a result, the clients' local models end up overfitting to their private datasets to a similar degree across all frameworks.

\begin{figure}[ht]
  \centering
  \begin{subfigure}[b]{0.32\linewidth}
    \includegraphics[width=\linewidth]{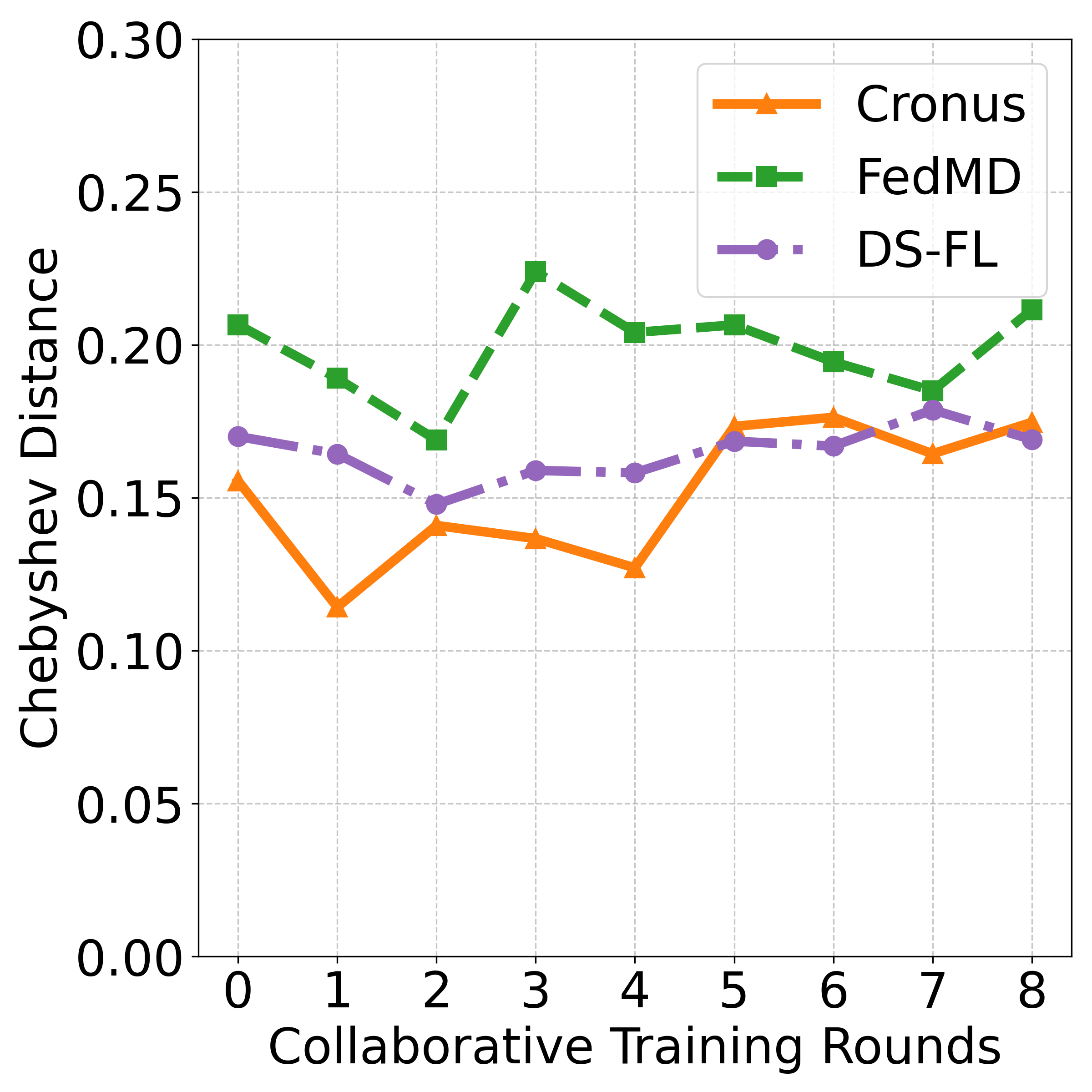}
    \caption{$\alpha=0.1$}
    \label{fig:ldia_result_diff_rounds_0.1}
  \end{subfigure}
  \hfill
  \begin{subfigure}[b]{0.32\linewidth}
    \includegraphics[width=\linewidth]{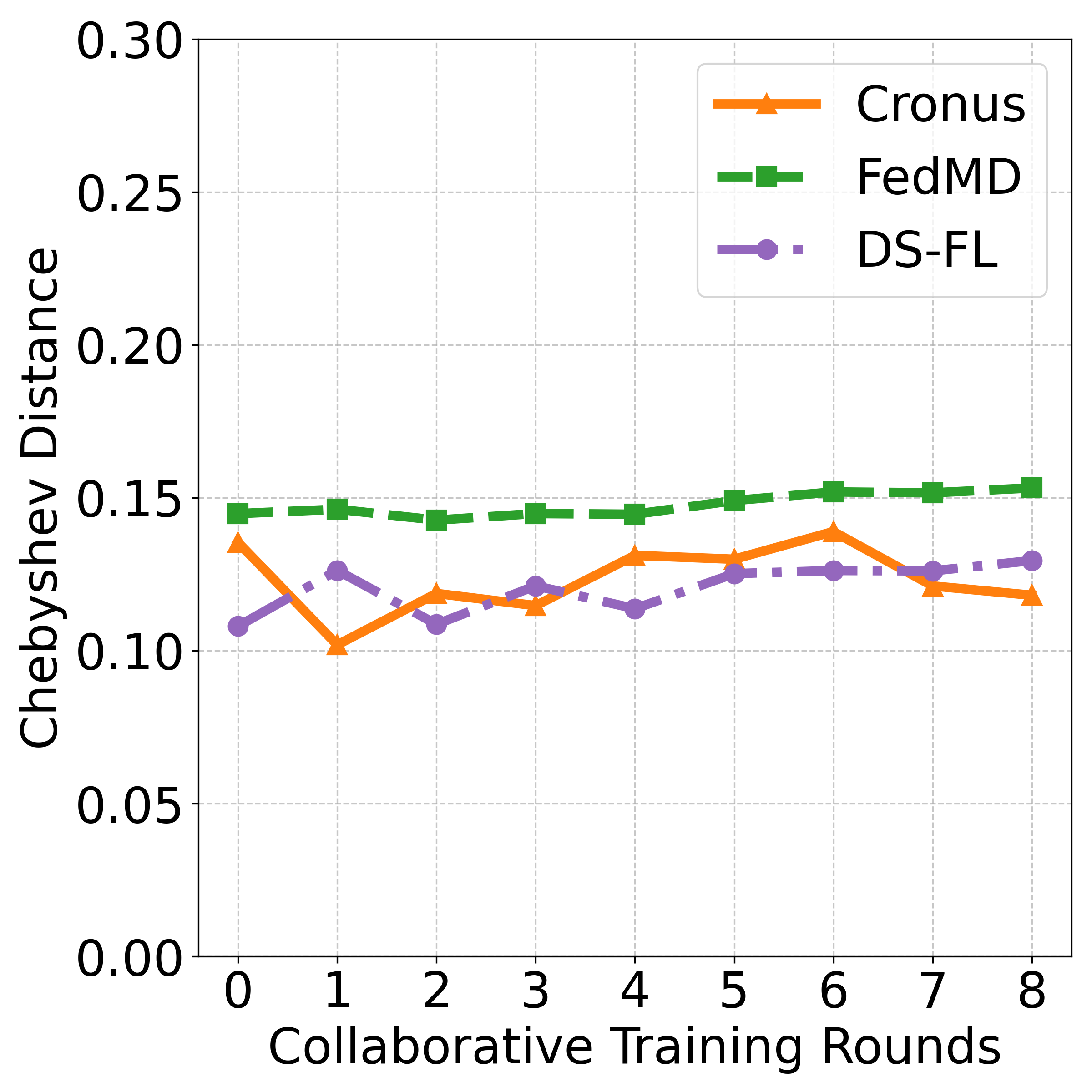}
    \caption{$\alpha=1$}
    \label{fig:ldia_result_diff_rounds_1}
  \end{subfigure}
  \hfill
  \begin{subfigure}[b]{0.32\linewidth}
    \includegraphics[width=\linewidth]{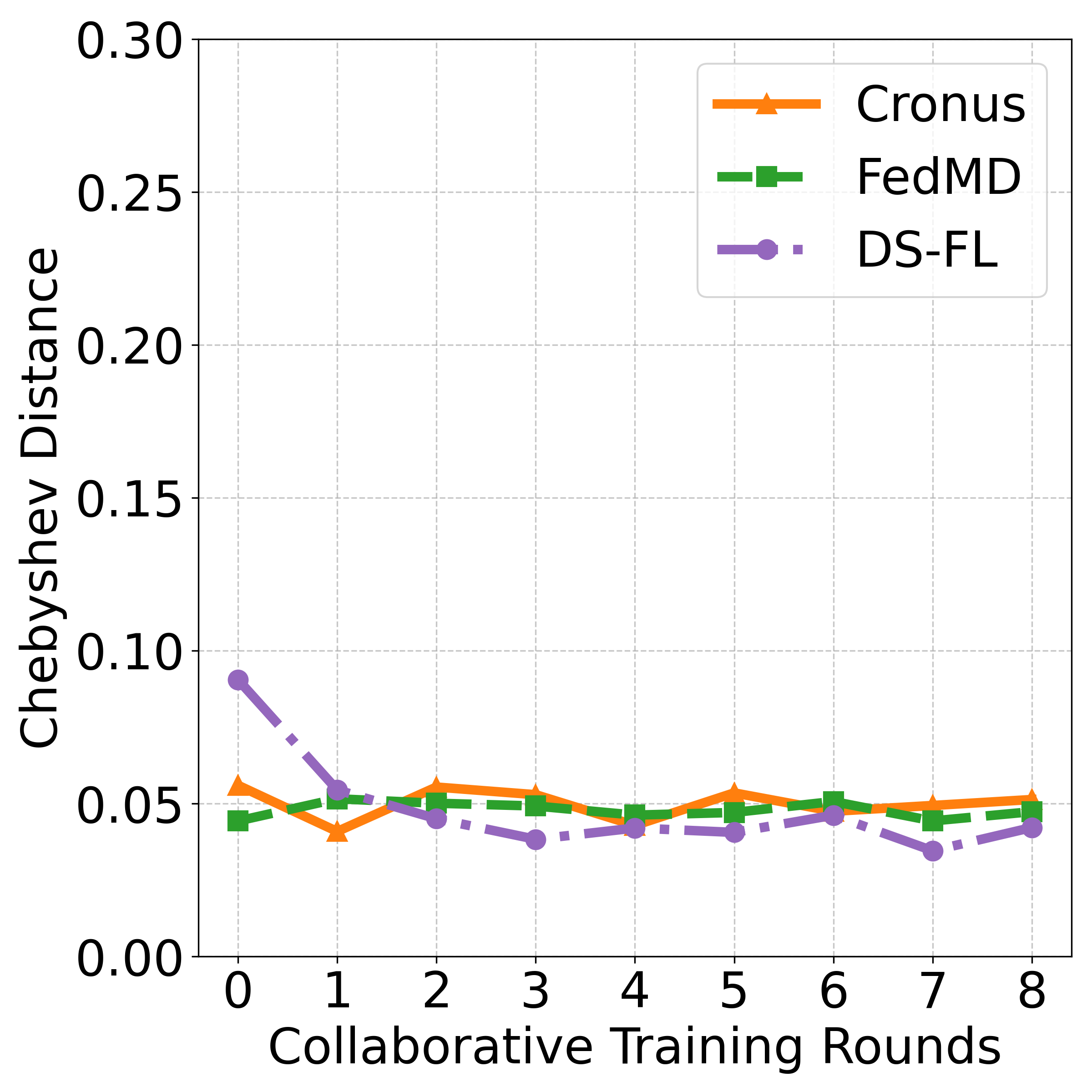}
    \caption{$\alpha=10$}
    \label{fig:ldia_result_diff_rounds_10}
  \end{subfigure}
    \caption{Chebyshev distance results of LDIA performed by the PDA-FD server on the CIFAR-10 dataset, shown for each collaborative training round.}
    \label{fig:ldia_result_diff_rounds}
\end{figure}

\textbf{Different Collaborative Training Rounds.} 
To evalutate the temporal dynamics of LDIA, we analyze its performance across multiple collaborative training rounds in various PDA-FD frameworks, as illustrated in Figure \ref{fig:ldia_result_diff_rounds}.
We aggregate the server's attack results across all clients for each round, to represent the overall LDIA performance over time.
The results reveal that the server successfully executes LDIA on clients in every round. 
Notably, the LDIA performance remains relatively stable as the number of collaborative training rounds increases, showing neither significant improvement nor decline.
This consistency can be attributed to the local updates phase preceding communication in each collaborative training round within PDA-FD frameworks. 
While this phase enhances knowledge transfer among clients, it simultaneously increases the degree of overfitting of each client's local model to their private data. 
This dual effect contributes to the observed stability in LDIA performance over multiple rounds.

\subsection{Experiment Results of MIAs}
\label{sec:MIA_eval}
In our MIA experiments, we evaluate the effectiveness of the proposed attack against 10 clients during the communication phase across three PDA-FD frameworks. 
Furthermore, we compare with existing attacks, including FD-leaks~\cite{yang2022fd}, MIA-FedDL~\cite{liu2023mia}, GradDiff~\cite{wang2024graddiff}, and Centralize-LiRA~\cite{carlini2022membership}, that serve as our baselines.

\textbf{Main Result.} We first evaluate co-op LiRA. 
Given that co-op LiRA is applicable in scenarios where clients' label distributions are similar, we conduct experiments across different datasets using a Dirichlet distribution with $\alpha=10$. This parameter setting ensures a more uniform distribution of labels across clients, aligning with co-op LiRA's operational scenario. 
Table \ref{tab:efficient_mia_main_result} presents the performance of co-op LiRA during the communication phase of the first
collaborative training round.
Our findings reveal that when the server attacks a specific client, utilizing only the other 9 clients' models as the reference models yields remarkably effective attack results.
This observation underscores the high efficiency and practicality of co-op LiRA, demonstrating its capability to achieve effective MIA without the need to train any additional reference models.

\begin{table*}[ht]
\caption{Performance of the server in conducting Co-op LiRA within the different PDA-FD Frameworks.}
\centering
\scriptsize
\resizebox{0.9\linewidth}{!}{%
\begin{tabular}{lccccccccccccc}
\toprule
\multirow{2}{*}{Datasets} &
\multicolumn{3}{c}{TPR at 1\% FPR} & 
\multicolumn{3}{c}{TPR at 0.1\% FPR} & 
\multicolumn{3}{c}{AUC} & 
\multicolumn{3}{c}{Balance Accuracy} \\ 
\cmidrule(lr){2-4}
\cmidrule(lr){5-7}
\cmidrule(lr){8-10}
\cmidrule(lr){11-13}
 &
 FedMD & DS-FL & Cronus & 
 FedMD & DS-FL & Cronus & 
 FedMD & DS-FL & Cronus & 
 FedMD & DS-FL & Cronus \\ 
 \midrule
 CIFAR10 
 & 22.15\% & 21.96\% &20.35\% 
 & 6.67\%  & 6.39\%  &5.76\% 
 & 0.819   & 0.850   &0.840 
 & 74.07\% & 77.11\% &76.23\% \\
\midrule
 CINIC10 
 & 10.97\% & 11.23\% & 10.99\%
 & 1.78\%  & 1.80\%  & 1.79\%
 & 0.794   & 0.811   & 0.815
 & 71.55\% & 73.89\% & 74.28\%\\
 \midrule
 Fashion-MNIST
 &3.24\% &1.80\% &1.64\%
 &1.09\% &0.31\% &0.33\%
 &0.582  &0.533  &0.531
 &55.89\%&55.38\%&54.28\% \\
\midrule
 Purchase 
 & 5.85\%  & 4.08\%  & 4.45\% 
 & 1.67\%  & 0.71\%  & 0.61\%
 & 0.616   & 0.712   & 0.709
 & 58.41\% & 66.46\% & 66.10\%\\
 \bottomrule
\end{tabular}%
}
\label{tab:efficient_mia_main_result}
\end{table*}

\begin{figure}[ht]
  \centering
  \begin{subfigure}[b]{0.325\linewidth}
    \includegraphics[width=\linewidth]{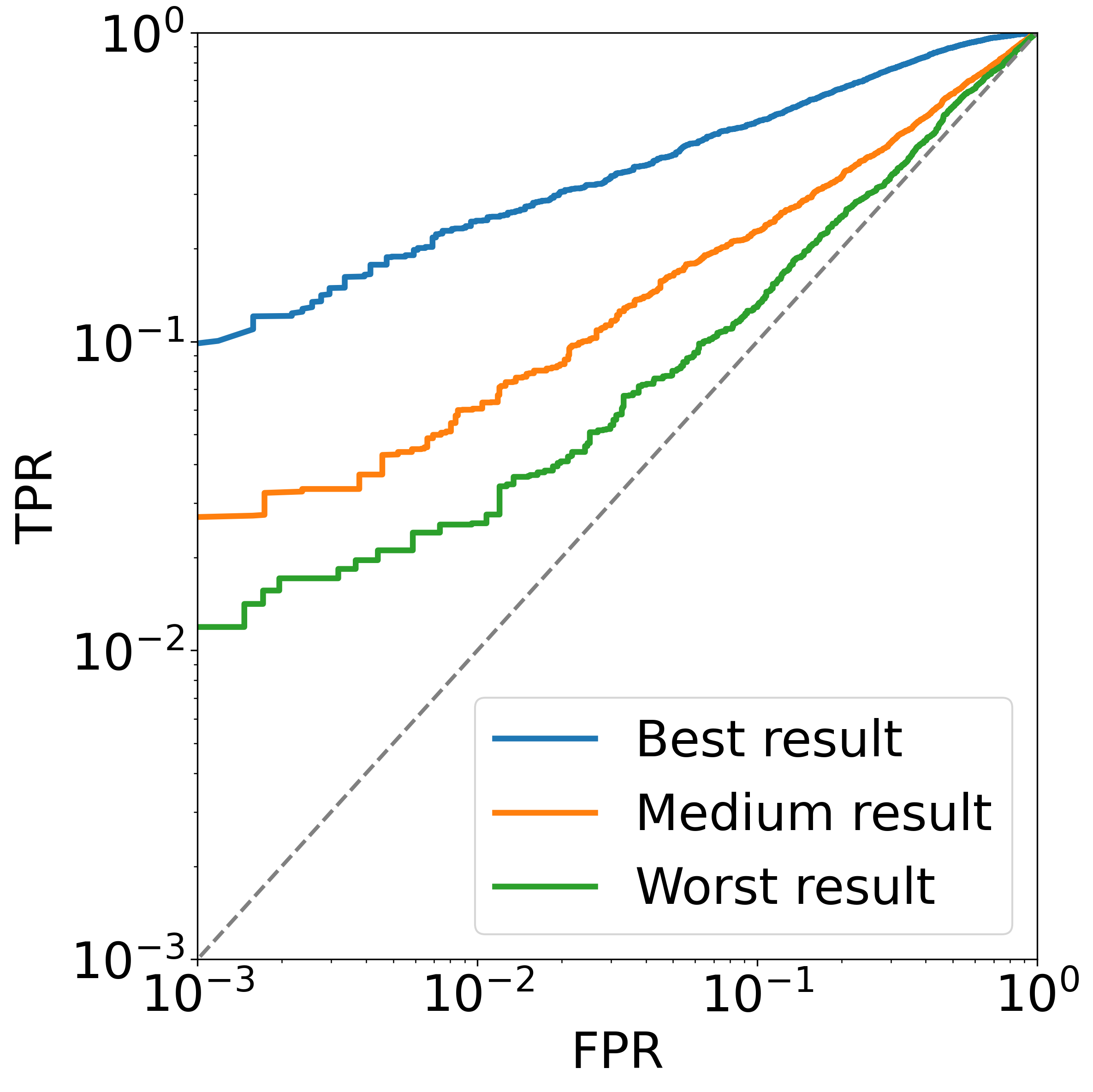}
    \caption{$\alpha=0.1$}
  \end{subfigure}
  \hfill
  \begin{subfigure}[b]{0.325\linewidth}
    \includegraphics[width=\linewidth]{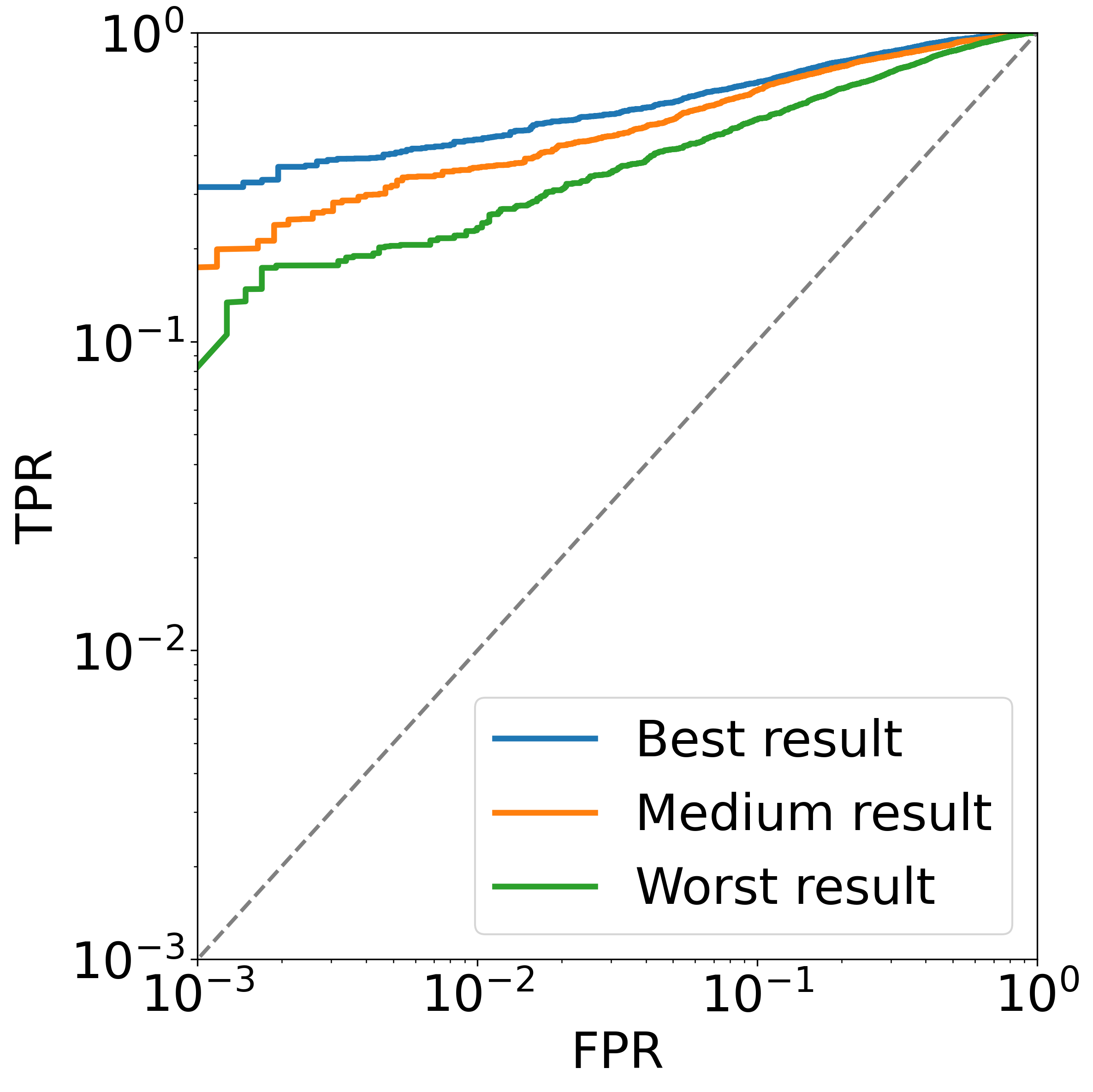}
    \caption{$\alpha=1$}
  \end{subfigure}
  \hfill
  \begin{subfigure}[b]{0.325\linewidth}
    \includegraphics[width=\linewidth]{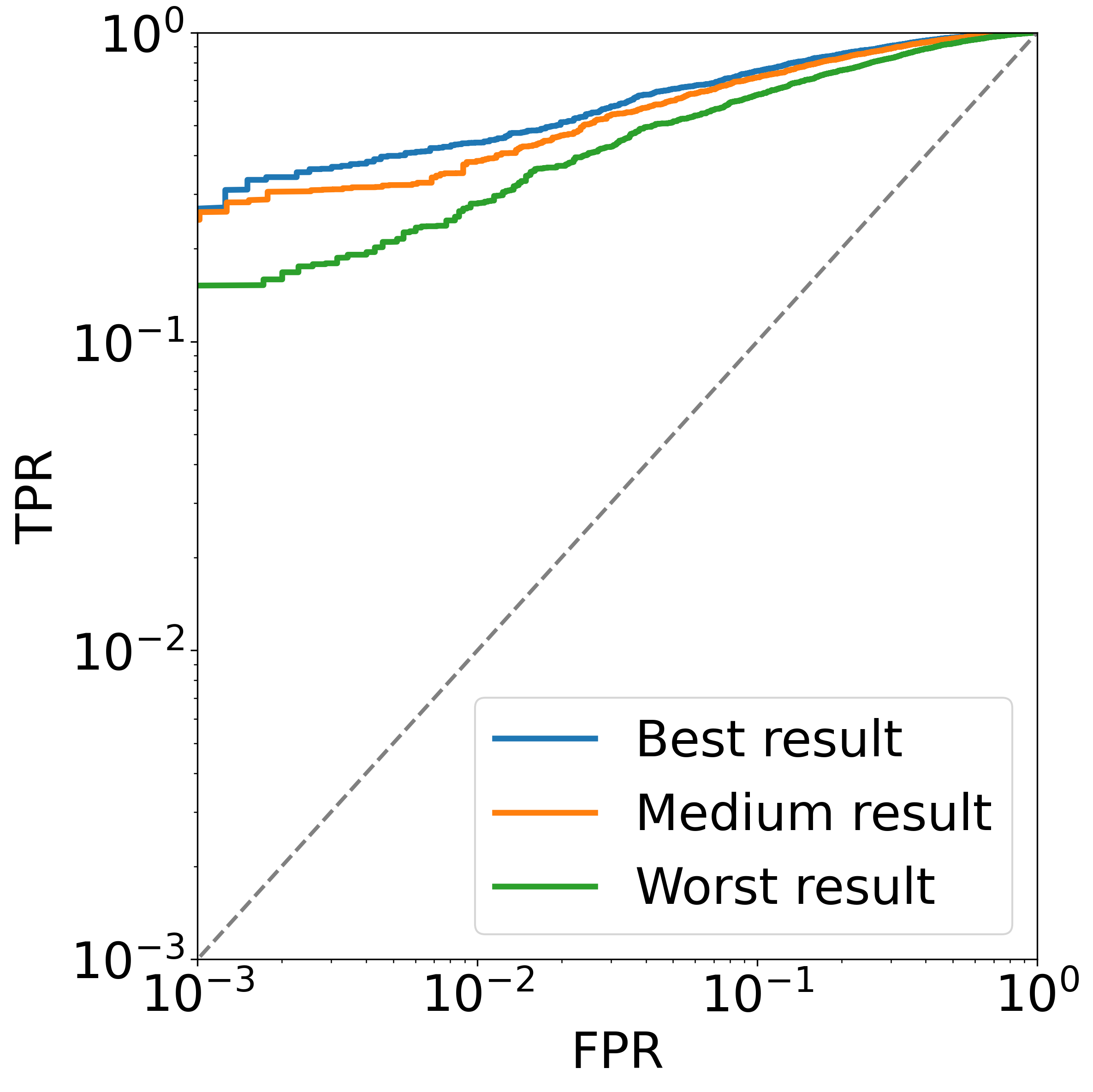}
    \caption{$\alpha=10$}
  \end{subfigure}
    \caption{Distillation-based LiRA performance of the DS-FL server on the CIFAR-10 dataset, presented as log-scale ROC curves under three distinct Dirichlet distributions.}
    \label{fig:distillation_mia_main_result}
\end{figure}

We subsequently evaluate the performance of distillation-based LiRA. 
For each client, we distill 32 reference models. 
The distillation dataset for each reference model consists of a randomly sampled 80\% subset of the public dataset used in the communication phase. 
This approach ensures a diverse set of reference models.
The model architecture of reference model is same as that of the target client model.
Table \ref{tab:distillation_mia_main_result} presents the average results of the server's MIA against all clients during the communication phase during the first collaborative training round. 
The results reveal that the server can launch highly effective MIA against the clients in the non-IID scenarios.
Figure \ref{fig:distillation_mia_main_result} presents the results of MIA experiments conducted in the DS-FL framework using the CIFAR-10 dataset, across various Dirichlet distributions.
Notably, we observe that even when the private dataset (CIFAR-10) and public dataset (CIFAR-100) have significantly different distributions, the server can still successfully launch MIA against clients by leveraging the distilled reference models from the public dataset. 
This finding underscores the efficacy of distillation-based LiRA in the PDA-FD frameworks, demonstrating its robustness to dataset disparities between the public and private data.

\begin{table*}[ht]
\caption{Performance of the server in conducting distillation-based LiRA within the different PDA-FD Frameworks.}
\centering
\scriptsize
\resizebox{0.9\linewidth}{!}{%
\begin{tabular}{lccccccccccccc}
\toprule
\multirow{2}{*}{Datasets} & 
\multirow{2}{*}{Setting} & 
\multicolumn{3}{c}{TPR at 1\% FPR} & 
\multicolumn{3}{c}{TPR at 0.1\% FPR} & 
\multicolumn{3}{c}{AUC} & 
\multicolumn{3}{c}{Balance Accuracy} \\ 
\cmidrule(lr){3-5}
\cmidrule(lr){6-8}
\cmidrule(lr){9-11}
\cmidrule(lr){12-14}
 & & 
 FedMD & DS-FL & Cronus & 
 FedMD & DS-FL & Cronus & 
 FedMD & DS-FL & Cronus & 
 FedMD & DS-FL & Cronus \\ 
 \midrule
 CIFAR10 
 & $\alpha$=10 
 & 20.94\% &35.76\% & 32.69\%
 & 10.42\% &19.42\% & 14.40\%
 & 0.764   &0.902   & 0.867
 & 69.68\% &82.01\% & 78.51\%\\
 & $\alpha$=1   
 & 17.62\% &29.28\% & 23.83\%
 & 8.11\%  &11.29\% & 5.77\%
 & 0.730   &0.839   & 0.804
 & 66.93\% &76.20\% & 72.94\%\\
 & $\alpha$=0.1 
 & 9.74\%  &12.89\% & 7.20\%
 & 2.09\%  &3.74\%  & 1.13\%
 & 0.618   &0.680   & 0.639
 & 58.91\% &63.49\% & 60.64\%\\
\midrule
 CIFAR10 
 & $\alpha$=10  
 & 11.20\% &34.61\% & 28.28\%
 & 1.48\%  &10.92\% & 5.49\%
 & 0.804   &0.901   & 0.891
 & 72.74\% &81.93\% & 80.97\%\\
 /CIFAR100 & $\alpha$=1 
 & 9.47\%  &25.94\% & 19.32\%
 & 0.93\%  &2.40\%  & 1.95\%
 & 0.758   &0.844   & 0.821
 & 68.92\% &76.61\% & 73.68\%\\
 & $\alpha$=0.1 
 & 7.79\%  &11.34\% & 5.61\%
 & 0.74\%  &0.51\%  & 0.88\%
 & 0.627   &0.686   & 0.652
 & 59.38\% &63.67\% & 61.75\%\\
\midrule
 CINIC10 
 & $\alpha$=10
 &13.83\% & 17.32\% & 15.91\%
 &3.12\%  & 4.15\%  & 3.85\%
 &0.741   & 0.855   & 0.834
 &71.42\% & 77.57\% & 75.26\% \\
 & $\alpha$=1 
 &10.96\% & 13.94\% & 12.07\%
 &2.37\%  & 3.59\%  & 3.01\%
 &0.704   & 0.781   & 0.757
 &68.93\% & 70.89\% & 69.21\%\\
 & $\alpha$=0.1 
 &5.91\%  & 6.81\% & 6.46\%
 &1.25\%  & 1.94\% & 1.72\%
 &0.649   & 0.652  & 0.661
 &60.27\% & 61.18\%& 62.59\% \\
 \midrule
 Fashion & $\alpha$=10  
 &3.07\% &1.85\%  & 1.73\%
 &0.91\% &0.35\%  & 0.21\%
 &0.588  &0.539   & 0.528
 &55.94\%&59.85\% & 55.43\% \\
 -MNIST  & $\alpha$=1  
 &3.30\% &1.71\% &  1.62\%
 &0.69\% &0.26\% &  0.25\%
 &0.583  &0.536  &  0.522
 &56.47\%&59.51\%&  54.31\%\\
 & $\alpha$=0.1 
 &1.88\% &1.39\% &  1.21\%
 &0.54\% &0.19\% &  0.23\%
 &0.538  &0.523  &  0.519
 &52.71\%&52.35\%&  51.32\%\\
\midrule
 Purchase 
 & $\alpha$=10 
 & 1.94\%   &5.91\%  & 2.43\%
 & 0.77\%   &1.31\%  & 1.93\%
 & 0.539    &0.665   & 0.706
 & 53.34\%  &62.69\% & 65.62\%\\
 & $\alpha$=1 
 & 1.98\%   &5.64\%  & 2.69\%
 & 0.83\%   &1.69\%  & 0.68\%
 & 0.534    &0.654   & 0.653
 & 53.31\%  &61.49\% & 62.24\%\\
 & $\alpha$=0.1 
 & 1.41\%  &5.04\% &  3.04\%
 & 0.42\%  &1.21\% &  1.19\%
 & 0.507   &0.591  &  0.588
 & 52.24\% &57.93\% & 57.69\%\\
 \bottomrule
\end{tabular}%
}
\label{tab:distillation_mia_main_result}
\end{table*}

\textbf{Different Data Distributions.}
We observe that the effectiveness of the proposed distillation-based LiRA attacks on clients decreases as the clients' label distributions become more imbalanced. 
This phenomenon can be explained by the fact that local models trained on datasets with highly skewed label distributions tend to produce disproportionately high posterior probabilities for the dominant labels.
This bias also affects the non-member samples that come from the same over-represented classes.
The core principle of LiRA relies on difficulty calibration, which becomes less effective in imbalanced scenarios.
As a result of this, the attacker's capability to discriminate between members and non-members is compromised.
This leads to a overall degradation in the performance of distillation-based LiRA on clients with highly imbalanced label distributions.

\textbf{Different PDA-FD Frameworks.}
The effectiveness of co-op LiRA remains relatively consistent across FedMD, DS-FL, and Cronus. 
However, for distillation-based LiRA, the effectiveness of MIAs ranks as follows: DS-FL achieves the best performance, followed by Cronus, with FedMD showing the least effectiveness.
In Cronus, clients upload softmax-processed posterior probability vectors rather than raw logits for public data during the communication phase. 
Compared to logits, the use of posterior probability vectors diminishes the server's ability to distill reference models that closely mimic the target model's performance.
Consequently, this limitation leads to a reduction in the effectiveness of MIA.
In the FedMD framework, clients train on public data before their private datasets during the local updates phase in their first collaborative training round. 
This process leads to clients training on all the target samples strategically selected by the server, regardless of their membership status.
While experiments demonstrate that subsequent training on private datasets reduces clients' memorization of public data, this initial exposure still impacts the server's MIA results.
However, when the public dataset is unlabeled, clients cannot train on it during the first collaborative training round. In this scenario, the server's MIA performance on clients remains unaffected.

\begin{figure}[ht]
\centering
  \resizebox{0.9\linewidth}{!}{%
  \begin{subfigure}[b]{0.49\linewidth}
    \includegraphics[width=\linewidth]{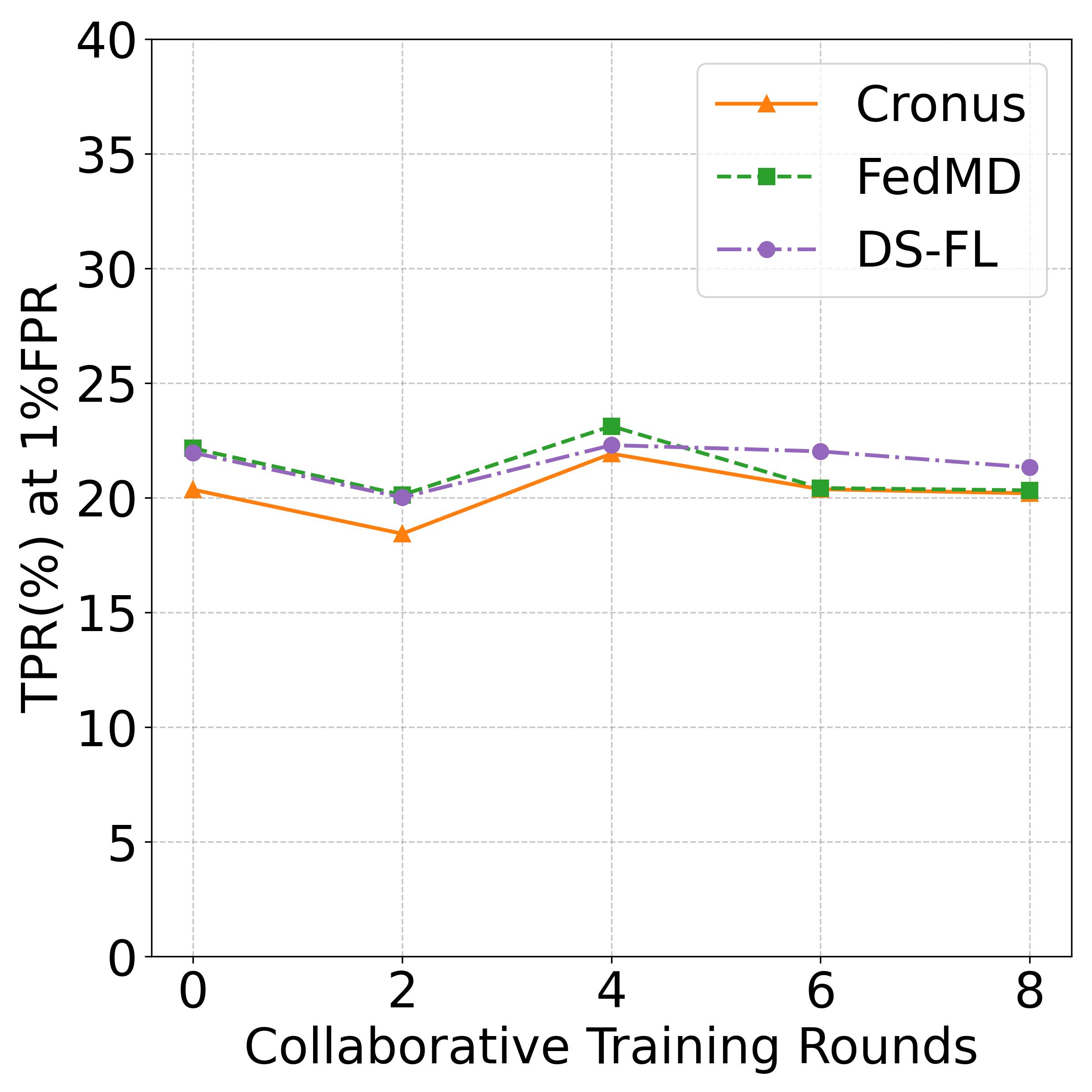}
    \caption{Co-op LiRA}
    \label{fig:mia_diff_round_1}
  \end{subfigure}
  \hfill
  \begin{subfigure}[b]{0.49\linewidth}
    \includegraphics[width=\linewidth]{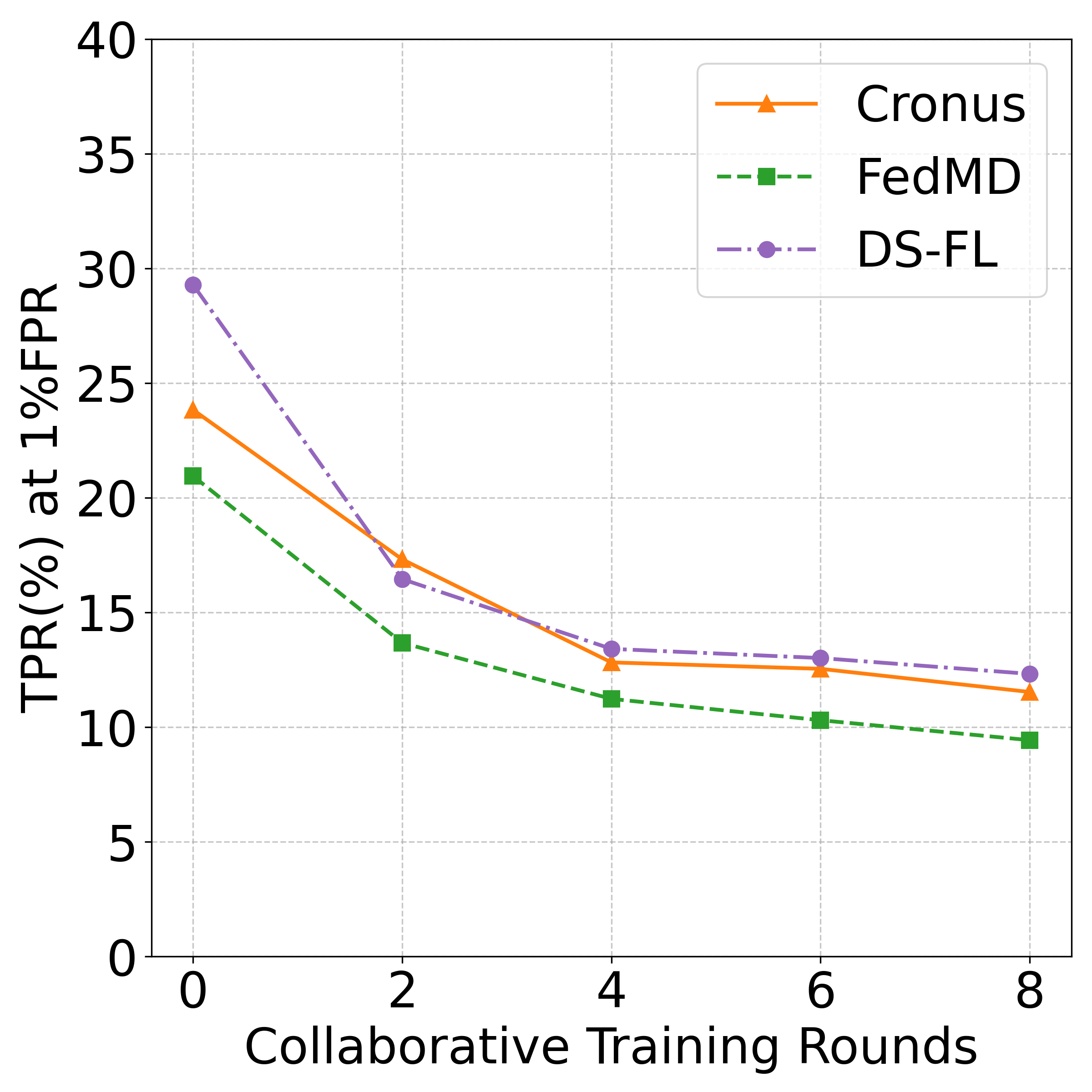}
    \caption{Distillation-based LiRA}
    \label{fig:mia_diff_round_2}
  \end{subfigure}
  }
    \caption{MIA performance across training rounds.}
    \label{fig:mia_diff_round}
\end{figure}

\textbf{Different Collaborative Training Rounds.}
Figure \ref{fig:mia_diff_round} illustrates the performance of the proposed MIAs in different PDA-FD frameworks across multiple collaborative training rounds on the CIFAR-10 dataset. 
To evaluate each MIA approach in its intended scenario, for co-op LiRA, we employ a Dirichlet distribution parameter $\alpha$=10.
While for distillation-based LiRA, we use $\alpha$=1. 
We use the average TPR at 1\% FPR of MIA across all clients as the metric to quantify performance.
The performance of distillation-based LiRA declines with more collaborative training rounds but eventually stabilizes.
We attribute this to the FD process, which gradually reduces the degree of overfitting of local models to their private data. 
While local updates maintain some private data memorization, the performance gap between local and distilled reference models for private data narrows over time.
The performance of co-op LiRA remains relatively stable as the collaborative training rounds increase. 
We attribute this stability to two key factors. 
First, the non-target clients in co-op LiRA cannot effectively learn membership information from the server-aggregated logits during the communication phase. 
Second, while the FD process reduces the overfitting level of local models to their private data, the local updates phase, where each client trains exclusively on its private dataset, maintains a consistent performance gap between local and reference models for their respective private data. 
This balance between reduced overfitting and continued exclusive training on private data likely contributes to the stability of co-op LiRA's performance across collaborative rounds.

\textbf{Comparison with Baselines.}
Our implementations of MIA-FedDL~\cite{liu2023mia} and GradDiff~\cite{wang2024graddiff} follow their proposed threat model settings, where the attacker can obtain a shadow dataset of data distribution consistent with the target model's training dataset to train the shadow model. 
For the implementations of FD-leaks~\cite{yang2022fd} and MIA-FedDL, we follow the same baseline experiment setup as described in GradDiff~\cite{wang2024graddiff} with one modification: we reposition the attacker from the client side to the server side. 
For Centralize LiRA~\cite{carlini2022membership}, we chose to implement a more realistic threat model where the server directly performs LiRA on the target client model, with shadow models obtained solely through direct training on the already available public dataset.
In our experiments, we chose to conduct experiments under the FedMD framework, with both the private dataset and the public dataset being CIFAR10. 
The parameter $\alpha$ of the Dirichlet distribution is set to 1.
The results are shown in Table~\ref{tab:baselines}.
\begin{table}[ht]
    \caption{Comparison with Baselines in FedMD(CIFAR10, $\alpha$=1).}
    \centering
    \begin{threeparttable}
    \scriptsize
    \resizebox{0.9\linewidth}{!}{%
    \begin{tabular}{c|c|c|c}
        \toprule
        Methods& TPR at 1\% FPR & AUC & Accuracy\\
        \midrule
        FD-leaks~\cite{yang2022fd} & 0.00\% & 0.552  & 54.34\% \\
        MIA-FedDL~\cite{liu2023mia}& 1.75\% & 0.643  & 59.31\% \\
        GradDiff~\cite{wang2024graddiff} & 3.04\% & 0.695  & 65.17\% \\
        Centralize LiRA~\cite{carlini2022membership} & 6.46\% & 0.701 & 62.40\%\\
        \midrule
        Co-op LiRA\tnote{*}  & 7.21\% & 0.711 & 64.48\%\\
        Distillation-based LiRA\tnote{*} & 17.62\% & 0.730 & 66.93\%\\
        \bottomrule
    \end{tabular}%
    }
    \begin{tablenotes}
    \scriptsize
      \item[*] Our attack
    \end{tablenotes}
    \end{threeparttable}
    \label{tab:baselines}
\end{table}

From the table, we can see that our proposed Co-op LiRA and Distillation-based LiRA exhibit stronger attack performance, especially on TPR at low FPR region, one of the most important metrics for measuring MIA performance~\cite{carlini2022membership}. FD-leaks shows limited attack performance due to low-quality shadow models. 
As for MIA-FedDL and GradDiff, even under stronger and overly idealistic threat model conditions, they still demonstrate weaker attack results compared to LiRA attacks because they rely on classifier-based MIA techniques instead of incorporating state-of-the-art MIA methods.
Meanwhile, our Co-op LiRA and Distillation-based LiRA outperform Centralized LiRA, further validating that our approaches for shadow model selection and training enable attackers to acquire higher-quality shadow models in realistic FD scenarios with constrained threat model assumptions, resulting in more effective attack results.
\section{Ablation Study}
\label{sec:ablation}
\subsection{Public Dataset Size}
The server in PDA-FD can control communication overhead by adjusting the size of the public dataset used in each collaborative training round.
We investigate its impact on the performance of LDIA and MIA. 
As public data does not affect co-op LiRA, our evaluations of MIA mainly focus on distillation-based LiRA. 
In our experiments, we use the DS-FL framework on the CIFAR-10 dataset with $\alpha=1$.
\begin{table}[h]
    \caption{Impact of Public Data Quantity on Label Distribution and Membership Information Leakage in PDA-FD.}
    \centering
    \scriptsize
    \resizebox{0.9\linewidth}{!}{%
    \begin{tabular}{c|c|c}
        \toprule
        Datasets size&  MIA (TPR at 1\%FPR) & LDIA (KL divergence)\\
        \midrule
        5000 & 29.28\% &  0.10  \\
        7500 & 31.84\% &  0.09 \\
        10000& 32.01\% &  0.07 \\
        \bottomrule
    \end{tabular}%
    }
    \label{tab:public_dataset_size}
\end{table}
Table \ref{tab:public_dataset_size} illustrates the degree of label distribution information and membership information leakage from clients when the quantity of the public data samples is set to 5000, 7500, and 10000, respectively.
The results indicate that larger public datasets contribute to increased privacy leakage risks for clients. 
We attribute this trend to two factors. For distillation-based LiRA, a larger public dataset provides a more extensive distillation dataset, enabling the attacker to obtain more robust reference models.
In the case of LDIA, a larger public dataset serving as the inference dataset allows the attacker to mitigate the impact of outliers or atypical data, thereby improving attack accuracy.

\subsection{Number of Reference Models}
\begin{figure}[h]
    \centering
    \includegraphics[width=0.9\linewidth]{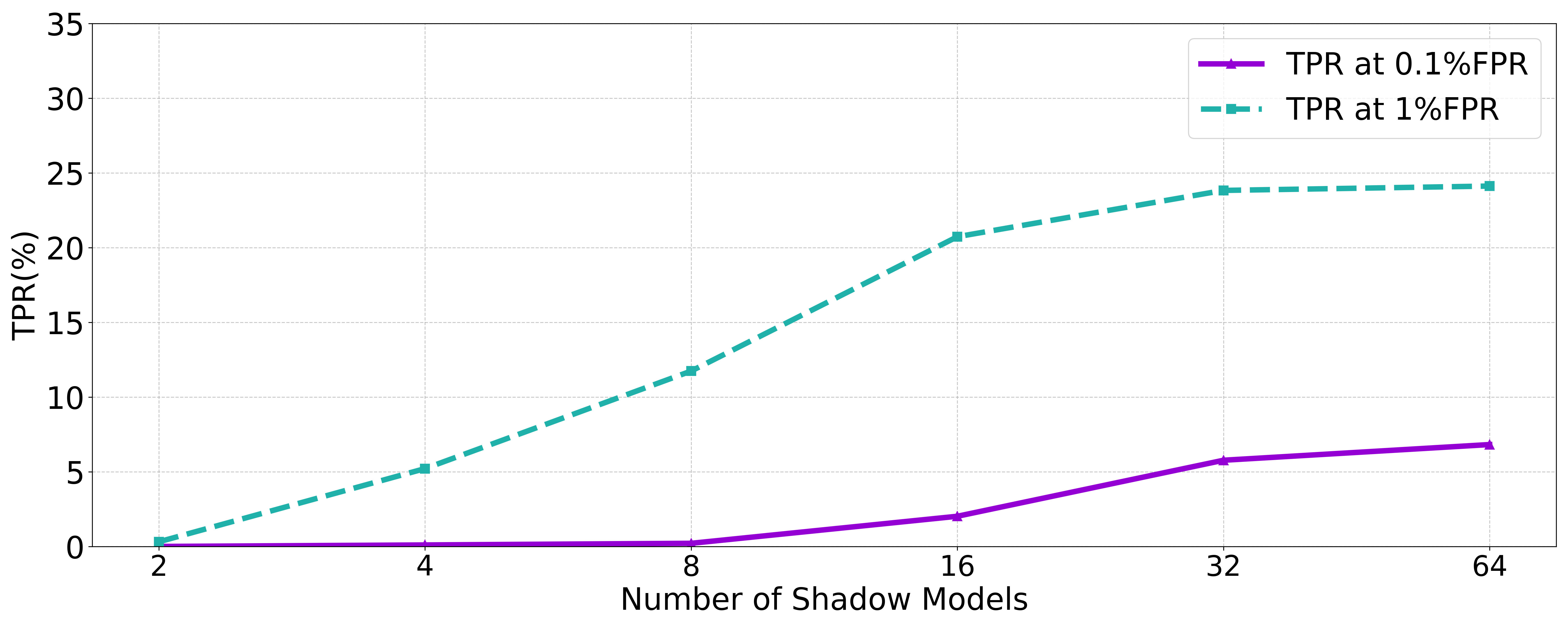}
    \caption{The performance of distillation-based LiRA vs. number of the distilled reference model.}
    \label{fig:distillation_lira_num_models}
\end{figure}

In LiRA, the attacker can form a more accurate Gaussian distribution by utilizing a larger number of reference models, thereby enhancing the precision of determining whether a target sample belongs to the target model's training data. 
We evaluate the performance of distillation-based LiRA with varying numbers of reference models. 
Figure \ref{fig:distillation_lira_num_models} shows results from experiments using the Cronus framework on CIFAR-10 with $\alpha=0.1$. 
The data reveals that the performance of the distillation-based LiRA's improves as the number of distilled reference models increases.

\subsection{Public-Private Data Distribution Shift}
We also evaluate the impact of varying degrees of data distribution shift between the public and private datasets on both our attacks and the FD training process.
We conduct our experiments following the FedMD federated distillation framework, where we set the Dirichlet distribution to $\alpha$ = 1 and the private dataset to CIFAR10. 
Then, we compare the attack efficacy of using different public datasets, including CIFAR10, CIFAR100, TinyImageNet~\cite{le2015tiny}, and SVHN~\cite{netzer2011reading}.
Table~\ref{tab:public_dataset_shift} shows the performance of FedMD and our attacks, LDIA and distillation-based LiRA, in different scenarios.
\begin{table}[h]
    \caption{Impact of Public-Private Data Distribution Shift.}
    \centering
    \scriptsize
    \resizebox{1\linewidth}{!}{%
    \begin{tabular}{c|c|c|c}
        \toprule
        Public Dataset&  MIA (TPR at 1\%FPR) & LDIA (KL divergence) & FedMD Accuracy \\ 
        \midrule
        CIFAR10 & 17.62\% & 0.17 & 75.38\% \\
        CIFAR100 & 9.47\% & 0.11 & 68.41\% \\
        TinyImageNet & 7.83\% & 0.11 & 60.01\% \\
        SVHN & 2.72\% & 0.18 & 47.12\% \\
        \bottomrule
    \end{tabular}%
    }
    \label{tab:public_dataset_shift}
\end{table}

We can see that as the data distribution shift between public and private datasets increases, both attacks (LDIA and Distillation-based LiRA) become less effective. 
For LDIA, larger distribution shifts make it harder for the inference dataset $D_{inf}$ in LDIA to reflect the target model's label sensitivity. 
For Distillation-based LiRA, increasing distribution shifts make it more difficult for shadow models to mimic target model performance by using knowledge distillation, reducing the attack effectiveness. 
Notably, we also find that larger distribution shifts simultaneously hinder the PDA-FD process itself, as effective knowledge distillation among clients becomes more difficult, ultimately degrading the overall federated learning performance.

\subsection{Resilience Against DP-SGD}
To evaluate the robustness of our proposed LDIA and MIA methods, we assess their effectiveness when the target client employs DP-SGD\cite{abadi2016deep} during the local updates phase. 
DP-SGD is a state-of-the-art privacy-preserving model training technique.
Our experimental setup includes 10 clients participating in DS-FL training on the CIFAR-10 dataset ($\alpha=10$). We conduct LDIA and Co-op LiRA attacks against the clients during the second round of training.
In DP-SGD, it introduces noise to gradients during training, governed by three key parameters. 
The clipping bound ($C$) limits the influence of individual data points on model parameters. 
The noise multiplier ($\sigma$) determines the amount of noise added to gradients. 
The privacy budget ($\varepsilon$) balances privacy guarantees and model utility, with smaller values providing stronger privacy at the cost of potentially noisier updates.
In our experiments, we set $C$ to be 10 and vary $\sigma$ to adjust $\varepsilon$. 
This setup allows us to evaluate our proposed attack under different privacy protection levels.
\begin{table}[h]
    \caption{Performance of MIA and LDIA against DP-SGD for DS-FL trained on CIFAR-10.}
    \scriptsize
    \resizebox{0.9\linewidth}{!}{%
    \centering
    \begin{tabular}{c|c|c|c|c}
    \toprule
         $\sigma$&$\varepsilon$&Average acc&LDIA(KL divergence)&MIA(TPR at 1\%FPR)\\
    \midrule
         0  &$\infty$ & 59.09\% & 0.03 & 15.76\% \\
         0.1& >10000  & 48.68\% & 0.07 & 2.86\% \\
         0.3& >5000   & 41.29\% & 0.08 & 2.11\% \\
         0.5& >2000   & 28.53\% & 0.09 & 1.54\% \\
         1.0& 231     & 21.34\% & 0.10 & 1.29\% \\
    \bottomrule
    \end{tabular}%
    }
    \label{tab:DP}
\end{table}

\subsection{Resilience Against Evasive Clients}
\label{sec:indirect_attacks}
To proactively protect their privacy,  cautious clients may choose to, in each communication round, avoid sending to the server the logits of some samples in the public dataset, particularly the ones that are also in its training dataset. 
To counter such defense, we propose two countermeasures as follows:
(1) In co-op LiRA, a shadow target model can be distilled using the logits of the samples in the public dataset provided by the target model, and then obtain the logits of the target sample from this shadow target model as an approximation to the one from the target clients' model. 
The intuition is that although knowledge distillation reduces the distilled student model's membership information of the teacher model~\cite{jagielski2024students}, it still preserves statistically significant enough membership information for a percentage of the members in teacher's training data, thus allowing some success in the MIA attack to the teacher model.
(2) We can also leverage a technique called indirect queries~\cite{wen2022canary, long2020pragmatic}, which is to first obtain logits of samples in the target sample's neighborhood from the target model and subsequently perform MIA using information encoded in these neighborhood logits. Neighbor samples are generated by adding noises to the target sample.

We conduct experiments to evaluate the effectiveness, and the experiments are on the CIFAR-10 dataset with a Dirichlet distribution parameter $\alpha$=10, with co-op LiRA as the MIA method.
Equipped with the first countermeasure, the attack achieves a TPR of 4.53\% at 1\% FPR.
Implementing a simplified version of the second countermeasure gives the attack a TPR of 4.23\% at 1\% FPR.
Note that in implementing countermeasure two, we add random Gaussian noise to the target samples to generate neighbor samples, with the noise clipped to the [-0.7, 0.7] range.
Studies in~\cite{wen2022canary, long2020pragmatic} implement more advanced schemes to learn from the neighbor logits, leading to better attacks. We leave studying such schemes as future work.

\section{Privacy Risk in Federated Distillation}
\label{sec:assessement}
While FL is designed to protect clients' private data, recent research~\cite{gu2023ldia,nasr2019comprehensive, liu2023mia,yang2022fd,wang2024graddiff} reveals significant privacy risks in these frameworks. 
In FL, Gu \etal~\cite{gu2023ldia} demonstrated that server-side LDIA could achieve a KL-divergence of 0.01 between the inferred and the ground truth label distributions on CIFAR-10. 
Nasr \etal~\cite{nasr2019comprehensive} showed that server or client-side MIA could reach accuracies of 92.1\% and 76.3\%, respectively, on CIFAR-100.

FD frameworks transfer distilled knowledge between participants instead of informative model parameters and gradients. This mechanism generally provides more privacy protection for each client's data than traditional FL frameworks (FedAVG, FedSGD, etc.).
However, through the lens of LDIA and MIA, we observe that although privacy leakage risk in FD appears less severe than in FL, significant risks remain, as state-of-the-art privacy attacks can still achieve non-trivial success rates based on the results in the literature and our experiments. 
Our work is the first to propose a LDIA method targeting the FD frameworks, and we achieve a KL divergence of 0.02 between the inferred and the ground-truth label distributions on CIFAR-10. This attack is less successful than in the traditional FL frameworks, but label distribution leakage has been demonstrated.
Targeting the PDA-FD frameworks,
Liu \etal~\cite{liu2023mia} proposed a client-side MIA method attaining 67.0\% balanced accuracy on CIFAR-100. 
Yang \etal~\cite{yang2022fd} also demonstrated a client-side MIA method that achieved an up to 75\% balanced accuracy on CIFAR-100.
Similarly, our MIA methods (co-op LiRA and distillation-based LiRA) demonstrate considerable server-side MIA effectiveness in achieving a TPR of up to 35.76\% at a 1\% FPR on CIFAR-10. 
In addition, effective MIA methods are reported to target other FD frameworks. 
For example, Wang \etal~\cite{wang2024graddiff} reported that their MIA attack achieved 67.06\% and 79.07\% accuracy on FedGen~\cite{zhu2021data} and FedDistill~\cite{jiang2020federated} respectively, on CIFAR-10.
One of the objectives of our study is to motivate future research on privacy risks in various FD frameworks and, more broadly, FL frameworks.

\section{Related Work}
\label{sec:related}

\BfPara{MIA and LDIA} 
Shokri \etal~\cite{shokri2017membership} pioneered MIA research by demonstrating how model output confidence scores could reveal training data membership. 
Nasr \etal~\cite{nasr2019comprehensive} extended this to FL, showing how both passive and active adversaries could exploit gradients and model updates.
LDIA represents another significant privacy threat in FL.
Gu \etal~\cite{gu2023ldia} introduced LDIA as a new attack vector where adversaries infer label distributions from model updates. Wainakh \etal~\cite{wainakh2021user} further explored user-level label leakage through gradient-based attacks in FL.
Recent works have exposed the vulnerability of FD to inference attacks.
Yang \etal~\cite{yang2022fd} proposed FD-Leaks for performing MIA in FD settings through logit analysis. Liu \etal~\cite{liu2023mia} and Wang \etal~\cite{wang2024graddiff} enhanced MIA using shadow models via respective approaches MIA-FedDL and GradDiff, though their assumptions were limited to homogeneous environments.

\BfPara{Defenses and Countermeasures}
DPSGD~\cite{abadi2016deep} can be employed during the training phase to mitigate against privacy attacks to the client model. 
Additionally, specialized MIA defense methods such as SELENA~\cite{tang2022mitigating}, HAMP~\cite{chen2023overconfidence} and DMP\cite{shejwalkar2021membership} can be integrated into the training process.
Several studies have proposed enhanced FD frameworks with improved privacy protection mechanisms to reduce client privacy leakage.
Zhu \etal~\cite{zhu2021data} investigated data-free knowledge distillation for heterogeneous federated learning.
They presented an approach that reduces the need for public datasets.
Chen \etal~\cite{chen2023best} proposed FedHKD, where clients share hyper-knowledge based on data representations from local datasets for federated distillation without requiring public datasets or models.


\section{Conclusion}
\label{sec:conclusion}
In this paper, we examine the privacy risk of using public datasets as the knowledge transfer medium in FD through the lens of label distribution information and membership information leakage, measured by attack success rates.
We evaluate three public-dataset-assisted FD frameworks (FedMD, DS-FL, and Cronus) using our proposed LDIA method and two MIA methods: Co-op LiRA and Distillation-based LiRA.
Our LDIA method performs considerably well measured in KL divergence (an average KL divergence of 0.10 in one setup), demonstrating a non-trivial risk level of label distribution information leakage. 
Co-op LiRA and Distillation-based LiRA, two MIA attacks for FD, achieve a state-of-the-art success rate evident by relatively high TPRs at low FPRs (up to 34.61\% TPR at 1\% FPR), indicating troublesome membership leakage risk.
These findings underscore the privacy vulnerabilities that persist in PDA-FD frameworks, highlighting the need for enhanced privacy-preserving mechanisms in FD environments.

\begin{acks}
We would like to thank the reviewers of PoPETs '25 for their invaluable feedback. 
This work is supported by an ONR grant N00014-23-1-2137.
\end{acks}

\bibliographystyle{ACM-Reference-Format}
\bibliography{reference}

\appendix

\section{Additional Experiment Details}

\subsection{Data Splits on Different Datasets}
\label{sec:datasets_division}
 The details of the specific partitioning of datasets(CIFAR10, CIFAR-10/CIFAR-100, CINIC-10, Fashion-MNIST and Purchase) and the number of classes used in our experiments.
\begin{table}[h]
    \caption{Datasets division.}
    \centering
    \resizebox{1\linewidth}{!}{%
    \begin{tabular}{c|c|c|c|c}
        \toprule
        Datasets&number of classes&$D_{train}$ & $D_{pub}$ & $D_{test}$\\
        \midrule
        CIFAR-10 & 10 & 40000 & 10000 & 10000 \\
        CIFAR-10/CIFAR-100 & 10 & 40000 & 10000 & 10000 \\
        CINIC-10 & 10 & 72000 & 18000 & 90000  \\
        Fashion-MNIST & 10 & 48000 & 12000 & 10000 \\
        Purchase & 10 & 21589 & 5397 & 11565 \\
        \bottomrule
    \end{tabular}%
    }
\end{table}

\section{Additional Ablation Study}
\subsection{Number of Epochs in Local Updates Phase}
Prior to the communication phase, clients train their local models on their private datasets during the local updates phase. 
This process enhances the local model's memorization of private data, facilitating knowledge transfer between clients but also potentially increasing privacy leakage. 
We measure the impact of the number of training epochs in the local updates phase on the leakage of label information and membership information from clients.
\begin{table}[h]
    \caption{Impact of Number of Training Epochs on Label Distribution and Membership Information Leakage in PDA-FD.}
    \centering
    \scriptsize
    \resizebox{0.9\linewidth}{!}{%
    \begin{tabular}{c|c|c}
        \toprule
        Number of Epochs&  MIA (TPR at 1\%FPR) & LDIA (KL divergence)\\
        \midrule
        2 & 8.10\% &  0.15  \\
        4 & 14.64\% &  0.10 \\
        6 & 15.43\% &  0.09 \\
        \bottomrule
    \end{tabular}%
    }
    \label{tab:epochs_local_updates}
\end{table}
As shown in Table \ref{tab:epochs_local_updates}, there is an increase in label distribution and membership information leakage from clients in DS-FL as the number of the local update training rounds increases from 2 to 6 on the CIFAR-10 dataset ($\alpha$=1).

\end{document}